Plamen Nikolov[a,b,c]* and Nusrat Jimi[a]

[a]*Department of Economics, State University of New York (Binghamton), Binghamton, NY USA*

[b]*Harvard University Institute for Quantitative Social Science, Cambridge, MA USA;*

[c]*IZA Institute of Labor Economics*

*corresponding author: Plamen Nikolov, Department of Economics, LT909, 3300 East Vestal Parkway, Binghamton, NY 13902

# What Factors Drive Individual Misperceptions of the Returns to Schooling in Tanzania? Some Lessons for Education Policy

Evidence on educational returns and the factors that determine the demand for schooling in developing countries is extremely scarce. Building on previous studies that show individuals underestimating the returns to schooling, we use two surveys from Tanzania to estimate both the actual and perceived schooling returns and subsequently examine what factors drive individual misperceptions regarding actual returns. Using ordinary least squares and instrumental variable methods, we find that each additional year of schooling in Tanzania increases earnings, on average, by 9 to 11 percent. We find that on average individuals underestimate returns to schooling by 74 to 79 percent and three factors are associated with these mis-perceptions: income, asset poverty and educational attainment. Shedding light on what factors relate to individual beliefs about educational returns can inform policy on how to structure effective interventions in order to correct individual misperceptions.

## 1. Introduction

Vast literature in the field of labour economics views education as an investment and posits that individuals decide to attain more schooling by comparing discounted future lifetime returns to schooling with the present costs for attendance. Measuring the future returns to schooling precisely, however, is fraught with econometric and data challenges (Card 2001) and is additionally complicated by the importance of survey designs (Serneels, Beegle and Dillon 2017), especially in developing countries.[1,2]  Furthermore, complex contextual and cognitive factors likely influence how individuals filter information about future educational returns and how they make actual decisions. Manski (1993) formalized the argument that subjective beliefs and future expectations about outcomes, including those to human capital investments and schooling, need to be taken into account so as to improve our ability to predict individual behaviour.

---

[1] Card (2001) and Duflo (2001) argue that returns to schooling are higher in developing countries than in high-income countries. However, evidence on the returns to schooling in developing countries is limited. Prior to Card (2001) and Duflo (2001), estimates on returns to schooling were largely based on Patrinos and Psacharopoulos (2010), who summarize the literature on the returns to education from low- and middle-income countries up to the 1990s.

[2] Using a survey experiment in Tanzania, Serneels, Beegle and Dillon (2017) investigate whether measurement error because of variation in survey design matters in estimating returns to education. The study shows that while the estimated returns to education differ according to the survey respondent they do not differ by the type of respondent.

In this paper, we use data from two surveys -- Tanzania's 2014 Integrated Labour Force Survey and a Dar es Salaam Perceived Returns Survey collected in 2014 – to estimate the returns to schooling and to measure individual beliefs about those schooling returns. Using ordinary least squares method (OLS) and instrumental variable (IV) approaches, we first estimate the returns to schooling at the primary and secondary-levels for Dar es Salaam, Tanzania. We then present data from the second survey that captures individual subjective perceptions on schooling returns in Dar es Salaam. In the final part of the analysis, we examine what factors drive the gap between the measured returns to schooling and the subjectively formed perceptions.

The possibility that perceptions, not measured returns based on earnings data, drive the individual demand for more schooling is particularly relevant and worrisome in light of low educational attainment rates in many developing countries -- particularly in Sub-Saharan Africa. Educational attainment in Sub-Saharan Africa, at the primary and secondary school levels, is extremely low. Among the 15 countries in Sub-Saharan Africa for which reliable educational data is available, only seven -- Kenya, Mauritius, Namibia, Seychelles, South Africa, Tanzania, and Zimbabwe -- have attainment rates for primary education at or above 50 percent (UNESCO 2011). Furthermore, the proportion of the population that completed at least a lower secondary school level ranges from 2 percent in Burkina Faso, to 70 percent in South Africa, with the median proportion of the population having attained a lower secondary school level of education of 21 percent. For example, in Tanzania, the percent having completed a lower secondary school level in 2011 is just 6 percent.

There are numerous supply and demand side factors that likely influence and constrain an individual in a developing country from obtaining more schooling. Supply side factors, for example, include distance to school, teacher training and the availability of textbooks and physical facilities. Demand side factors comprise enrolment fees, uniforms, the quality of the educational experience and the opportunity cost of one's time spent in school.

Because schooling generates important monetary benefits, in some settings, it may be more appropriate to examine the role individual perceptions of these monetary benefits play and their interaction with one's perceived opportunity cost of schooling, if any, in influencing one's educational demand. If individuals misperceive the future monetary benefits from obtaining more schooling, correcting such misperceptions may be a very cost-effective approach towards increasing school participation. Jensen (2010) examines whether providing additional information regarding the monetary returns to students can affect subsequent enrolment in a study with 8th-graders in the Dominican Republic. Students at a randomly selected subset of schools were informed of the returns estimated from earnings data. Relative to students not provided with this information, students in the treated schools were nearly 4 percentage points (7 percent) more likely to be enrolled in school the next academic term, and 4 years later had completed on average about 0.20 more years of schooling. The striking findings of the study demonstrate that careful information targeting can have powerful influence on behaviour change and that it can be a very cost-effective strategy for improving educational outcomes, especially in low-income contexts.

Using ordinary least squares and instrumental variable estimations, we find that the average returns are commensurate with the returns to schooling in developed countries in other studies (Card 2001). Using data from individuals in the Dar es Salaam area, we measure the perceived average earnings for males who are 40 to 50 years old with primary schooling only and secondary schooling only. We also obtain the actual average measured earnings for males of the same age group who completed the two educational levels and live in the same geographic area. We find that, on average, individuals underestimate the actual average earnings approximately by 74 to 79 percent. Finally, we examine what factors are associated with the gap between the measured earnings and the subjectively perceived average earnings and we find that three main factors drive that gap: low earnings, asset poverty and educational attainment. The lowest earners and the lowest decile in asset poverty is the group that underestimates educational returns the most.

Our findings show that individuals substantially underestimate the returns to primary and secondary schooling and we point to those demographic groups that underestimate the educational returns the most. This kind of information can be an important input to policy-makers in designing effective information targeting interventions. However, such approach merits an important caution for policy makers. Although information targeting may be a relatively inexpensive approach to boost the demand for schooling in the short run, it is important to consider the potential general equilibrium effects of such a policy and its dynamic impact on the returns to education, especially for the segments of the labour market that become more abundant.

This paper contributes to the existing labour economics literature in three major ways. First, it provides estimates of the returns to education from a representative household survey in a Sub-Saharan context, where rigorous evidence of educational returns is lacking, and more evidence is important because educational attainment remains persistently low. Second, the paper provides evidence that individuals underestimate earnings for individuals with primary schooling by 74 percent and underestimate earnings for individuals who finish secondary schooling by 79 percent. Although numerous factors likely constrain an individual's decision to obtain more schooling, our paper provides evidence that there is scope for an information provision policy intervention that can accurately refocus individual misperceptions regarding the monetary returns to education. In fact, previous studies in Latin America document that such interventions can result in large improvements in educational outcomes (Jensen 2010). Finally, we provide evidence on the socioeconomic characteristics of individuals who misperceive educational returns the most. The lowest earners and the lowest decile in asset poverty are the groups that underestimate educational returns. This finding complements the findings by Hedges et al. (2016) who show that in Tanzania, overall wealthier families are more likely to invest in education.

The remainder of the paper is structured as follows. We present the survey data used in Section 2. We present the methodology used in Sections 3 and 4. In Section 5, we present the results. We conclude in Section 6.

## 2. Educational System in Tanzania

The educational system in Tanzania follows a 2-7-4-2-3-plus structure (see Table 1). The first two years comprise pre-primary schooling, followed by seven years of primary school, four years of ordinary secondary school (called ordinary level), two years of advanced secondary school (called advanced level) and at least three years of higher education. The main language of instruction for the primary-level is Kiswahili. At the secondary and higher education levels, the language of instruction is English. Only primary school (i.e., the first seven years beyond the pre-primary-level) is compulsory. Upon primary school completion, students take a national exam, a pre-requisite for transitioning onto the secondary-level. Approximately 50 percent of students transitioning from the primary education level pass that test (Development Partners Group in Tanzania 2013).

**Table 1. Tanzanian Educational System by Years of Schooling.**

| Years of Schooling | Diplomas/Degrees in Tanzania |
|---|---|
| 7 | Primary (Standard 1-VII) |
| 11 | Secondary ordinary level (Form 1-4) |
| 13 | Secondary advanced level (Form 5 and 6) |
| 16 or more | University education |

Secondary school comprises grades 8-13. Upon completion of the ordinary secondary school level (i.e., grade 11), students take national exams that cover 10 subjects. Approximately 18 percent of students pass the secondary school exams to move onto Form 5 (grades 12 and 13).

Secondary schools cover the following subjects: agriculture, commerce, home economics and technology. Students in forms 1 through 4 (ordinary level) study civics, English, Kiswahili, history, geography, physics, chemistry, biology and mathematics. Forms 5 and 6 (advanced level) cover commerce, arts and social sciences, natural sciences and general studies. Students who successfully obtain the Advanced Certificate of Secondary Education are eligible to apply for admission to institutions of higher education.

## 3. Survey Data

### 3.1 The 2014 Integrated Labour Force Survey

The 2014 Integrated Labour Force Survey (ILFS) in Tanzania is intended to track labour-related trends in the country. The survey sample was designed to provide labour market information in three main areas: Dar es Salaam, other urban areas and rural areas. Its sampling frame is based on a 2012 Census. The survey sample followed a three-step process. The first step involved sampling of enumeration areas (EA) within each stratum from the ordered list of EAs in the sampling frame. Four-hundred-and-eighty EAs were selected during this first step -- 360 urban EAs and 120 rural EAs. The second step involved sampling households from each of the selected EAs. Twenty-four households were sampled from each selected EA. The third step entailed selecting individual respondents. Prior to the data collection, a household roster was prepared for individuals aged five years or above. From this roster and based on the Kish selection grid method, a household member was identified for the time-use module and a survey was administered. Table 2 reports the summary statistic of the resulting sample.

<div align="center"><strong>Table 2. Summary Statistics.</strong></div>

| | Tanzania Mainland | Dar es Salaam |
|---|---|---|
| Gender (1 if male) | 0.47 | 0.47 |
| | (0.50) | (0.50) |
| Age (in years) | 34.59 | 33.36 |
| | (15.37) | (13.72) |
| Schooling (in years) | 8.51 | 10.45 |
| | (5.04) | (4.59) |
| Monthly Earnings (in Tanzanian Shillings) | 144,824.00 | 194,559.20 |
| | (350,729.00) | (438,272.10) |
| Experience[a] | 20.07 | 17.33 |
| | (15.51) | (15.55) |
| Never Attended School (1 if schooling is zero) | 0.10 | 0.04 |
| | (0.30) | (0.19) |
| Attained Primary Education Only (1 if schooling>=7) | 0.75 | 0.86 |
| | (0.43) | (0.35) |
| Attained Secondary Ordinary Level Education (1 if schooling>=11) | 0.18 | 0.26 |
| | (0.38) | (0.44) |
| Attained Secondary Advanced Level Education (1 if schooling>=13) | 0.07 | 0.11 |
| | (0.26) | (0.32) |
| Observations | 29,035 | 12,310 |

*Note*: Data source is Tanzania's 2014 Integrated Labor Force Survey. .[a]Experience is calculated by taking the difference of one's age and one's schooling minus six years following Mincer (1974), Boissiere, Knight and Sabot (1985) and Lemieux (2006). Primary, Ordinary Secondary, and Advanced Secondary education refers to individuals who completed only primary, ordinary secondary, and advanced secondary schooling, respectively. Standard deviations in parentheses.

Earnings and labour force variables, the focus of this paper, were collected in the labour force module. The labour force module is comprised of two forms: the Labour Force Survey Form 1 (LFS1) and the Labour Force Survey Form 2 (LFS2). LFS1 was administered to a household head or a knowledgeable representative. The first form was designed to capture information on household characteristics such as number of household members, disability, migration, level of education, training, household economic activities, household amenities, access to public services and asset ownership. LFS2, on the other hand, was focused primarily on individual-level data. The form was an individual questionnaire administered to individuals aged five or above. It covered labour force-related information, such as past, current, primary and secondary economic activities, unemployment, hours of work and income streams from various employment activities. Using data from the ILFS's LFS2 form, we primarily drew on data for three variables: earnings, schooling and individual socioeconomic characteristics.

### 3.2 Dar es Salaam Perceived Returns Survey

The second survey was conducted in Dar es Salaam through the months of July to September of 2014. The sampling was designed to produce a sample that when appropriately weighted is representative of households in metropolitan Dar es Salaam at the time of the survey, including both households with and without young adult residents. Approximately 1,300 households were selected to answer a questionnaire that included socioeconomic questions. The survey proceeded in a stratified two-step sample selection process. The first step entailed the selection of sample clusters. The second step entailed the selection of households within each cluster.

The first step entailed selecting sample clusters, based on the probability proportional to size (PPS) approach, using the 2002 Tanzania Enumeration Areas (EAs). To account for the measure of size we used the number of households in each EA, as measured by the 2002 Population Census. Following this method provided for efficiency in obtaining equal sub-sample sizes across the two-step selection. In the second step, households were chosen from each of the selected clusters.

The survey questionnaire collected information on education, employment, earnings, demographic and socioeconomic characteristics for all household members. In addition to collecting data for these questions, the head of the household was asked to provide a subjective assessment of the earnings of current 40 to 50 years old individuals with two levels of education:

> Now, we would like you to think about adult men who are about 40 to 50 years old and who have completed only [primary school/secondary school]. Think not just about the ones you know personally, but all people like this throughout the country. How much do you think they earn in a typical week, month or year?

This survey question attempts to measure the respondent's perceptions regarding individual earnings for an individual who has attained a given educational level. It is asked in a hypothetical third person in order to purge the question from one's own beliefs about himself/herself or factors such as own tribe or own ethnicity in the subjective elicitation about average earnings. Therefore, the question can be used to assess whether perceptions about the returns to schooling differ from measured returns because of inaccurate information on prevailing wages in the labour market.

This simple question, however, has several downsides. First, the question does not specify the precise meaning of "expected" earnings nor does it address aspects of earnings that deal with other properties of the earnings distribution—such as the mean or median. The question also does not factor in future uncertainty, life-course profile of earnings, or the individual expectation regarding inflation.

This elicitation approach follows Nguyen (2008), a study implemented in Madagascar whose survey instrument was designed for the context of a developing country. Attanasio and Kaufmann (2009) and Kaufmann (2014) employ a different approach that captures individual beliefs more precisely. However, the instruments developed by Attanasio and Kaufmann (2009) and Kaufmann (2014) rely on significantly more complicated questions that deal with abstract, hypothetical situations stated in formal and complicated language.[3]

## 4.      Empirical Methodology

### 4.1 Age-Earnings Profiles

Based on the sample from Tanzania's 2014 Integrated Labour Force Survey, we construct the age-earnings profiles, which show the mean measured earnings at various ages. We construct these profiles for two main purposes. First, we take advantage of the age-earnings profiles to compute the average measured earnings by various educational levels. We use these data in in our subsequent analysis where we compare the average objective earnings to the average perceived earnings. Second, we construct our own age earnings profiles to depict how individual earnings change over the life cycle.

We estimate the profile from a cross-sectional regression of earnings on age, encompassing all full-time individual workers aged 15 years and above in the sample year:

$$Earnings_i = \sum_{n=1}^{N} \alpha_n \, Age_i^n + e_i$$

_______________________

[3] Attanasio and Kaufmann (2009) and Kaufmann (2015) ask what individuals expect is the maximum and minimum they might earn under different education scenarios, as well as the probability of earning more than the midpoint of these two. With an assumption on the distribution of expectations, the data can be used to estimate various moments of the distribution.

where Earnings$_i$ represents the earnings of individual i and $Age_i^n$ is a dummy variable taking a value of 1 if individual i is in age category n and 0 otherwise.

### 4.2 Returns to Schooling in Tanzania: OLS Estimates

Following Heckman, Lochner et al. (2006) and the standard approach used within the returns to schooling literature, we estimate a Mincerian wage equation:

$$\ln(Y_i) = \alpha_i + \beta_1 S_{j,i} + \beta_3 E_i + \beta_4 E_i^2 + \beta_5 Gender_i + \varepsilon_i \qquad (1)$$

where $Y_i$ denotes earnings, $S_{j,i}$ denotes individual's school completion of level $j$, $E_i$ denotes experience[4], and $Gender_i$ denotes one's gender (1 if male).[5] We proxy the declining returns to experience with a quadratic term $E_i^2$. Using this specification, we examine the wage impacts of two schooling levels – primary and secondary. The coefficient of interest is $\beta_1$, which describes the percent change in earnings from having attained a respective schooling level – either primary or secondary.

Estimates on the OLS method, however, are unlikely to isolate the true causal effects as the method is ill-equipped to separate the wage effects of schooling from the wage effects of other hard-to-measure factors (Angrist and Krueger 1999, Card 2001, Griliches 1977). The empirical literature with regard to the direction of potential bias stemming from OLS estimation is ambiguous. The schooling coefficient is likely afflicted by two competing sources of bias: positive ability and comparative advantage bias as well as attenuating measurement error bias. Because the measurement error inherent in documenting educational attainment is mean-regressive –i.e., individuals with the highest level of schooling cannot report positive errors and those with the lowest level of schooling cannot report negative errors – most of the previous studies concur on likely positive bias of the OLS estimates. We attempt to improve the accuracy of the OLS estimates by employing an instrumental variable approach described in the next section.

---

[4] We create a proxy variable for experience by computing the difference between one's age and one's schooling minus six years. This approach is standard in the literature based on Mincer (1974), Boissiere, Knight and Sabot (1985) and Lemieux (2006).
[5] We follow the approach proposed by Becker (1964) and Psacharopoulos (1994, p 1326) who argue that the inclusion of too many control variables can artificially cause downward bias in the returns to education. Pereira and Martins (2004) address this issue in detail.

### 4.3 Return to Schooling: Instrumental Variable Estimation

A large number of studies have attempted to employ an instrumental variable approach to estimate the returns to schooling. A valid instrumental variable must meet two conditions: relevance and exogeneity. The relevance condition requires that the instrument be correlated with the number of years of schooling that an individual attains. The exogeneity condition requires that the instrument affects earnings only through the endogenous schooling variable. Previous studies explore a range of potential instruments, including changes in schooling laws (Harmon and Walker 1995), proximity to college (Card 1993) and birth quarters (Angrist and Krueger 1991).[6] Following (Angrist and Krueger 1991), we use data on birth quarters from the ILFS survey to estimate (2) below by instrumenting $S_{j,i}$ with instrument $Q1_i$ -$Q3_i$, denoting individual birth quarters with specification (3):

$$\ln(Y_i) = \alpha_i + \beta_1 S_{j,i} + \beta_2 E_i + \beta_3 E_i^2 + \beta_4 Gender_i + \varepsilon_i \qquad (2)$$

$$S_{j,i} = \pi_{1,1} Q1_i + \pi_{1,2} Q2_i + \pi_{1,3} Q3_i + x_i' \pi_{1,x} + v_{1,i} \qquad (3)$$

Q1-Q3 in (3) denote individual birth quarters. The compulsory schooling law in Tanzania is based on age, not number of years of school and therefore we can expect that people born at different times of the year can drop out after receiving different amounts of schooling. We examine the correlation between the proposed instruments (Q1-Q3), which are used as a source of identifying variation for the endogenous variable on schooling ($S_{j,i}$). As stated previously, the exclusion restriction for the validity of Q1-Q3 as instruments requires no direct relationship between the quarter of birth and earnings. In other words, the identifying assumption is that the timing of a person's birth is unrelated to inherent traits (e.g. motivation, intelligence). Thus, the birth timing should not have a direct effect on wages, but only affect wages through the relationship with completed schooling induced by compulsory education laws.[7]

---

[6] Rather remarkably, most IV estimates appear to be larger than the corresponding OLS estimates, suggesting that OLS may underestimate the true returns to education. One potential explanation could be the measurement error for measuring schooling outcomes. Alternatively, the returns to education are highly heterogeneous and the parameters identified by the IV strategy are local average treatment effects (LATE), which describe the returns to education only for the subsample in which the IV induces more years of schooling (Imbens and Angrist 1994).

[7] However, if the instrument is valid, IV produces local average treatment effects (LATE), which may reflect non-representative effects with heterogeneous cost or return functions (Card 2001).

# 5. Results

## 5.1 Age Earnings Profiles

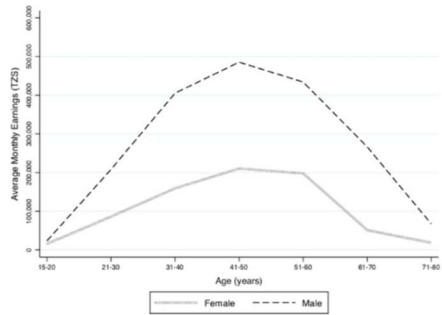

**Figure 1. Age-Earning Profile by Gender.**

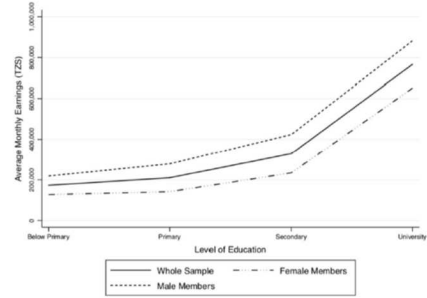

**Figure 2. Average Monthly Earnings by Educational Level in Dar es Salaa**

*Note:* Data source: Tanzania's 2014 Integrated Labor Force Survey. Sample includes employed males and females of Dar es S. 15 to 80. Monthly earnings are adjusted for inflation with 2012 serving as the base year. Exchange rate in 2012: $1=1,615 TZS

*Note:* Data source: Tanzania's 2014 Integrated Labor Force Survey. Sample includes employed males and females of Dar es Salaa to 80. Monthly earnings are adjusted for inflation with 2012 serving as the base year. Exchange rate in 2012: $1=1,615 TZS. 'Belov to people who attended school but did not finish primary-level education; 'Primary' refers to people who completed primary-level 'Secondary' refers to people who completed secondary-level schooling only; University refers to people who attended university-

**Table 3. Average Monthly Earnings by Age Groups.**

| | Age 15-20 | Age 21-30 | Age 31-40 | Age 41-50 | Age 51-60 | Age 61-70 | Age 71-80 | All |
|---|---|---|---|---|---|---|---|---|
| **Panel A: Females** | | | | | | | | |
| Primary | 18,683.86 | 68,188.83 | 104,129.88 | 147,471.66 | 102,632.92 | 53,899.50 | 62,758.63 | 79,344.76 |
| | (41,858.88) | (196,026.94) | (188,906.61) | (276,276.25) | (192,171.26) | (147,940.39) | (266,262.35) | (191,033.11) |
| Observations | 814 | 1,238 | 1,016 | 521 | 183 | 57 | 18 | 3,847 |
| Secondary | 7,280.82 | 126,427.38 | 311,815.44 | 455,421.64 | 528,553.68 | 90,751.71 | 0 | 192,627.35 |
| | (30,489.88) | (271,493.99) | (595,796.18) | (706,792.44) | (654,093.37) | (162,659.21) | 0 | (450,112.67) |
| Observations | 416 | 769 | 443 | 177 | 92 | 16 | 3 | 1,916 |
| **Panel B: Males** | | | | | | | | |
| Primary | 25,516.70 | 214,185.38 | 302,631.32 | 309,331.75 | 296,971.38 | 157,006.59 | 105,566.07 | 225,920.98 |
| | (68,185.33) | (364,101.70) | (441,000.87) | (443,512.18) | (337,091.49) | (298,189.96) | (186,150.64) | (380,111.02) |
| Observations | 572 | 789 | 893 | 545 | 248 | 90 | 26 | 3,163 |
| Secondary | 19,378.77 | 205,212.04 | 591,608.95 | 834,350.32 | 721,002.73 | 515,147.78 | 90,833.60 | 404,949.23 |
| | (73,046.45) | (413,648.00) | (727,351.32) | (1,081,174.80) | (839,980.20) | (1,165,151.60) | (187,188.42) | (724,476.06) |
| Observations | 375 | 759 | 555 | 313 | 168 | 69 | 8 | 2,247 |
| **Panel C: All** | | | | | | | | |
| Primary | 21503.76 | 125,017.28 | 196,985.39 | 230,223.77 | 214,456.44 | 117,026.29 | 88,053.93 | 146,555.88 |
| | (54,374.76) | (283,008.66) | (346,005.31) | (379,863.73) | (300,251.38) | (255,223.16) | (220,518.33) | (311,037.62) |
| Observations | 1,386 | 2,027 | 1,909 | 1,066 | 431 | 147 | 44 | 7,010 |
| Secondary | 13,020.08 | 165,561.90 | 471,787.33 | 697,472.00 | 652,905.37 | 435,261.46 | 66,060.80 | 307,229.14 |
| | (55,235.78) | (351,511.68) | (696,443.93) | (979,195.78) | (783,430.24) | (1,063,750.80) | (162,258.48) | (622,625.38) |
| Observations | 791 | 1,528 | 998 | 490 | 260 | 85 | 11 | 4,163 |

*Note:* Data source: Tanzania's 2014 Integrated Labor Force Survey. The sample includes employed males and females of Dar es Salaam aged 15 to 80 years. Monthly earnings are adjusted for inflation with 2012 serving as the base year (Exchange rate in 2012: $1=1,615 TZS). Primary School Completion and Secondary School Completion covers people who passed primary and secondary school respectively. Standard deviations in parentheses.

Based on data from the ILFS survey, Figure 1 displays the age-earnings profiles for males and females, ages 15-80 (Table 3 reports the data for these groups). Figure 1 exhibits several earnings patterns. First, earnings in Tanzania rise until the late 30s and early 40s and then level off, a pattern that seems to mirror the path of earnings in many other developed countries, except for the fact that earnings' tapering in Tanzania occurs much earlier.[8] Second, the earnings peak for both genders occurs approximately halfway through the span of one's working life. Third, the age-earnings profile exhibits a typical concave shape akin to the profiles in other developed countries (Polachek and Siebert 1993, p. 16). Lastly, the most rapid earnings growth occurs early in one's career.

The age-earnings profiles also exhibit a few stark differences by gender. First, the female earnings are lower than the male earnings. Male earnings peak at approximately age 40, while female earnings peak at around age 51. While the age-earnings profiles are concave for both genders, the initial growth of earnings for males occurs at a much faster pace than it does for females. Furthermore, the female wage profiles are flatter than they are for males as male wage profiles exhibit a precipitous decline in later life. This pattern can be accounted for by worker expectations regarding future discontinuity in labour force participation.[9]

Interpreting differences in earnings for various age groups based on age-earnings profiles is fraught with difficulties related to disentangling cohort effects from age and period effects. While diminished physical vigour and mental alertness, the obsolescence of education and skills, or the decision to work shorter hours partially accounts for declining incomes of older workers, an even stronger force behind the decline relates to the cross-sectional character of the data.[10,11]

---

[8] Based on data from developed countries, Reynolds, Masters, and Moser (1987, p. 91) argue that "a person's earnings normally rise with age until somewhere beyond the age of 40, then level off, and eventually decline."

[9] Polachek (1975) and Weiss (1981) argue that a worker who anticipates discontinuous labor force participation acquires on-the-job training at a different rate than the worker who anticipates continuous employment.[9] Women's earnings mirror the pattern of workers who anticipate discontinuous labor force participation and their absence from the labor market is generally due to childbearing. Therefore, women's earnings exhibit a flatter and often non-monotonic pattern (their age-earnings profiles exhibit a midlife dip) depending on the degree of intermittent work behavior (Mincer and Polachek 1974).

[10] Rodgers et al. (1996), for example, identify several cohort-related influences, including the size of a cohort (labor supply vs demand), varying rates of inflation, and varying rates of productivity growth, both for individuals and the economy. The cohort problem can be especially pronounced for women.

[11] In the U.S., women entering the labor market towards the conclusion of the 20th-century had greater access to education, greater access to professions, higher societal acceptance of women in the workforce, and faced less of a starting wage differential than women in cohorts entering before them. These cohort effects can potentially skew a typical cross-sectional age-earnings profile. Rodgers et al. (1996) explores age-earnings profiles along racial, gender, and educational lines. In their empirical analysis, they show that there is a correlation between more education and higher earnings expectations. Once again, if women in younger cohorts are taking advantage of educational opportunities at a greater rate than women in older cohorts, this should impact their access to the labor market and their earnings.

### 5.2 OLS-based Returns to Schooling in Tanzania

We first present the results based on the Mincerian wage specification. Tables 4 and 5 display the results.

**Table 4. Returns to Education.**

| | Log of Monthly Earnings (TZS) | | | |
| --- | --- | --- | --- | --- |
| | Tanzania mainland | | Dar es Salaam | |
| | (1) | (2) | (3) | (4) |
| Experience[a] | 0.05*** | 0.05*** | 0.05*** | 0.05*** |
| | (0.00) | (0.00) | (0.00) | (0.00) |
| Experience Squared | -0.00*** | -0.00*** | -0.00*** | -0.00*** |
| | (0.00) | (0.00) | (0.00) | (0.00) |
| Gender (1 if male) | 0.57*** | 0.56*** | 0.60*** | 0.59*** |
| | (0.02) | (0.02) | (0.03) | (0.02) |
| Primary (1 if schooling≥7) | 0.44*** | 0.44*** | 0.31*** | 0.32*** |
| | (0.03) | (0.03) | (0.06) | (0.05) |
| Ordinary Secondary (1 if schooling≥11) | | 1.26*** | | 0.98*** |
| | | (0.04) | | (0.05) |
| Advanced Secondary (1 if schooling≥13)[b] | 1.56*** | 2.11*** | 1.29*** | 1.88*** |
| | (0.04) | (0.04) | (0.06) | (0.06) |
| Constant | 10.16*** | 10.17*** | 10.51*** | 10.48*** |
| | (0.04) | (0.03) | (0.06) | (0.06) |
| $R^2$ | 0.24 | 0.27 | 0.25 | 0.30 |
| Mean Log Monthly Earnings (TZS) | 11.71 | | 12.09 | |
| Observations | 16,817 | 16,817 | 7,256 | 7,256 |

*Note*: Tanzania's 2014 Integrated Labor Force Survey. Monthly earnings are adjusted for inflation with 2012 serving as the base year (Exchange rate in 2012: $1=1,615 TZS). Standard errors in parentheses. *** Significant at the 1 percent level. ** Significant at the 5 percent level. * Significant at the 10 percent level.[a]Experience is calculated by taking the difference of one's age and one's schooling minus six years following Mincer (1974), Boissiere, Knight and Sabot (1985) and Lemieux (2006). [b] In column 1 and column 3, education is divided into two groups: Primary and Secondary. Primary and Secondary refers to individuals who completed only primary and only secondary schooling, respectively. Therefore, Advanced Secondary schooling in columns 1 and 3 actually implies secondary education (people with ordinary secondary or above schooling). In columns 2 and 4, Primary, Ordinary Secondary, and Advanced Secondary refers to individuals who only completed primary, ordinary secondary, and advanced secondary schooling, respectively.

Columns (1)-(2) display the results for mainland Tanzania, whereas Columns (3)-(4) show the results from the sample in Dar es Salaam. The results clearly show that human capital acquisition in the form of more schooling is associated with higher earnings. Most notable is the pattern of coefficient magnitudes for each level of schooling. Estimated returns to each level of education in Table 4 demonstrate that returns grow by 44 percent for primary school to 126 percent for ordinary secondary school and to 211 percent for advanced secondary school. The implied return for an additional year of schooling is between 6 to 16 percent per annum, all else equal. In contrast, previous studies detect higher returns for primary-level education in developing countries (Psacharopoulos and Patrinos 2002).[12] Columns (3)-(4) of Table 4 show the returns for primary school completion, ordinary secondary school level and the advanced secondary-levels for Dar es Salaam only. The implied educational returns are 32 percent for primary schooling, 98 percent for ordinary secondary school and 188 percent for advanced secondary school, respectively. The implied return for an additional year of schooling in Dar es Salaam is between 5 and 14 percent per annum, all else equal.

**Table 5. Returns to Education: By Gender.**

| | Log of Monthly Earnings (TZS) | | | |
| --- | --- | --- | --- | --- |
| | Tanzania mainland | | Dar es Salaam | |
| | (1) | (2) | (3) | (4) |
| Experience[a] | 0.05*** | 0.05*** | 0.05*** | 0.05*** |
| | (0.00) | (0.00) | (0.00) | (0.00) |
| Experience Squared | -0.00*** | -0.00*** | -0.00*** | -0.00*** |
| | (0.00) | (0.00) | (0.00) | (0.00) |
| Gender (1 if male) | 0.59*** | 0.59*** | 0.64*** | 0.62*** |
| | (0.04) | (0.06) | (0.19) | (0.09) |
| Primary (1 if schooling≥7) | 0.41*** | 0.41*** | 0.27*** | 0.27*** |
| | (0.04) | (0.05) | (0.07) | (0.07) |
| Ordinary Secondary (1 if schooling≥11) | | 1.37*** | | 1.07*** |
| | | (0.06) | | (0.08) |
| Advanced Secondary (1 if schooling≥13)[b] | 1.71*** | 2.39*** | 1.43*** | 2.16*** |
| | (0.06) | (0.05) | (0.07) | (0.08) |
| Gender Dummy * Primary (1 if schooling≥7) | 0.05 | 0.05 | 0.06 | 0.07 |
| | (0.05) | (0.05) | (0.09) | (0.19) |
| Gender Dummy * Ordinary Secondary (1 if schooling≥11) | | -0.18* | | -0.15 |
| | | (0.06) | | (0.10) |
| Gender Dummy * Advanced Secondary (1 if schooling≥13) | -0.24** | -0.45*** | -0.24** | -0.45*** |
| | (0.07) | (0.07) | (0.10) | (0.11) |
| Constant | 10.14*** | 10.15*** | 10.49*** | 10.46*** |
| | (0.05) | (0.05) | (0.07) | (0.07) |
| $R^2$ | 0.24 | 0.27 | 0.26 | 0.30 |
| Mean Log Monthly Earnings (TZS) | 11.71 | | 12.09 | |
| Observations | 16,817 | 16,817 | 7,256 | 7,256 |

*Note:* Tanzania's 2014 Integrated Labor Force Survey. Monthly earnings are adjusted for inflation with 2012 serving as the base year (Exchange rate in 2012: $1=1,615 TZS). Standard errors in parentheses. *** Significant at the 1 percent level. ** Significant at the 5 percent level.* Significant at the 10 percent level. [a]Experience is calculated by taking the difference of one's age and one's schooling minus six years following Mincer (1974), Boissiere, Knight and Sabot (1985) and Lemieux (2006).. [b]In column 1 and column 3, education is divided into two groups: Primary and Secondary. Primary and Secondary refer to individuals who completed only primary and secondary schooling, respectively. Therefore, Advanced Secondary schooling in columns 1 and 3 actually implies secondary education (i.e., people with ordinary secondary or above schooling-levels). In columns 2 and 4, Primary, Ordinary Secondary, and Advanced Secondary refer to people who completed only primary, ordinary secondary, and advanced secondary schooling, respectively. Standard errors in parentheses.

Tables 5 displays the results based on the Mincerian wage equation estimation by gender. Table 5 clearly shows that males in Tanzania earn between 59 percent and 62 percent more than females, all else equal. Columns (1)-(2) display the results for mainland Tanzania; Columns (3)-

(4) show the results from the Dar es Salaam sample. The results show that we do not detect gender differences in earnings for individuals who only complete primary schooling. However, at the advanced secondary-level, we do observe gender playing an important role in shaping the relationship between educational attainment and earnings. Returns to education at the advanced secondary level are higher for females than their male counterparts. In both mainland Tanzania and in Dar es Salaam, the return is on average 45 percent lower for males than females. In both cases, this implies approximately 3.5 percent lower return for an additional year of schooling for male compared to female.

Our result that the returns to schooling are higher for females than males has been previously documented in studies using earnings data from Africa (Psacharopoulos & Patrinos, 2004). The debate on the causes of these higher returns for women has yet to be settled in the context of Africa or elsewhere (Dougherty, 2005). Numerous explanations for the returns differential by gender, however, have been put forward. Among the conjectures, the most frequent are: greater ability bias for women (Deolalikar 1993), lower educational attainment for women than for men (Schultz 2002), occupational segregation by gender (Deolalikar 1993; Dougherty 2005), sample selection, gender differences in traits (Deolalikar 1993), male-differential in the quality of education attained (Dougherty 2005), and factors related to discrimination, tastes, and circumstances (Dougherty 2005).

Given the limitations of the ILFS dataset and that the gender differential is not a primary objective of this paper, we only provide suggestive evidence on a few factors influencing the differential schooling returns in the Tanzanian context. In particular, we look at three potential factors: average educational attainment by gender, gender-based occupational segregation and gender differences in the number of hours worked. First, we examine the average educational attainment by gender (Appendix A Table A1 reports the results). The table shows that females, on average, have lower educational attainment than males. In particular, Columns (3) and (4) of Appendix Table A1 show that the male-female difference between their average educational attainments is statistically significant. Second, another possible explanation of the differential in the male-female schooling coefficients relates to a sector-occupation composition effect -- females could be underrepresented in jobs or sectors where schooling is a relatively unimportant factor in the determination of earnings. We provide suggestive evidence that the composition effect plays a role in three ways. We first document in Appendix Table A5 and Appendix Table

A8 the average gender composition in major employment sectors and employment occupations. To examine whether the monetary returns to schooling are mediated by sectoral and occupational sorting by gender, we include sectoral controls (in Appendix Table A6) and occupational controls (in Appendix Table A9) in the Mincerian specifications. If, indeed, the wage effects of education are induced by changes in sector-occupation selection, we would observe the estimated effects of educational levels to change[13] once we account for the job-sectoral controls. The coefficients associated with the three educational levels indeed change (generally shrink) once we include sectoral and occupational controls, which is suggestive that some of the male-female differential is mediated by sectoral-occupational sorting by gender.[14] Furthermore, we also examine, based on Dougherty (2005), how the male-female differential changes once we account for sample selection by interacting gender, years of schooling and sectoral dummies (in Appendix Table A7) as well as the male-female differential in returns to schooling within the sectors/occupation (in Appendix Table A10). The triple interactions in both tables (by sector in Appendix Table A7 and by occupation in Appendix Table A10) are generally statistically significant, which is suggestive evidence that sectoral and occupational segregation plays an important role in driving the education premium by gender. Finally, we examine the role of the number of hours worked as another potential factor to account for gender differences in the rate of return to schooling. We mimic the analytical approach described above and present the results in Appendix Tables A2, A3 and A4. The results show that on average, women work fewer hours than men do, and that the number of hours worked is also an important determinant of the male-female differential for the rate of return to schooling.

### 5.3 Instrumental Variables Estimates of the Return to Schooling

We then turn our attention to the OLS-IV estimation comparing the returns to schooling for individuals who finished only primary-level schooling with individuals who finished secondary-level schooling. Table 6 presents the OLS results based on specification (1) in columns

---

[13] The direction of coefficient change will depend on whether more educated men sort positively or negatively into higher paying occupations.
[14] These coefficients should not be interpreted as causal effects since the specifications do not address the endogeneity of sectoral and occupational choice.

1-4. The IV estimates based on specifications (2)-(3) are presented in columns 5-8 for mainland Tanzania and Dar es Salaam.

**Table 6. Returns to Education: Instrumental Variable Estimates.**

| Variable | Log of Monthly Earnings (TZS) | | | | | | | |
|---|---|---|---|---|---|---|---|---|
| | OLS estimates | | | | IV estimates | | | |
| | Tanzania mainland (1) | Tanzania mainland (2) | Dar es Salaam (3) | Dar es Salaam (4) | Tanzania mainland (5) | Tanzania mainland (6) | Dar es Salaam (7) | Dar es Salaam (8) |
| **Panel A: Primary Level** | | | | | | | | |
| Primary | 0.51*** | 0.45*** | 0.35*** | 0.32*** | -0.74 | 1.14 | 0.97 | 2.93 |
| | (0.02) | (0.03) | (0.03) | (0.05) | (1.42) | (1.42) | (1.41) | (2.49) |
| Constant | 11.89*** | 10.14*** | 11.44*** | 10.38*** | 12.16*** | 9.47*** | 10.95*** | 8.07*** |
| | (0.02) | (0.03) | (0.05) | (0.04) | (0.36) | (0.15) | (1.27) | (2.22) |
| Controls | - | yes | - | yes | - | yes | - | yes |
| Observations | 12,857 | 12,857 | 4,901 | 4,901 | 9,801 | 9,801 | 4,324 | 4,324 |
| $R^2$ | 0.03 | 0.14 | 0.01 | 0.17 | 0.00 | 0.14 | 0.00 | 0.00 |
| **Panel B: Ordinary Secondary Level** | | | | | | | | |
| Ordinary Secondary | 0.76*** | 0.90*** | 0.56*** | 0.71*** | -0.35 | 0.76 | 0.13 | 0.88* |
| | (0.02) | (0.02) | (0.03) | (0.03) | (0.51) | (0.51) | (0.51) | (0.46) |
| Constant | 11.61*** | 11.00*** | 12.01*** | 11.29*** | 11.85*** | 10.86*** | 11.92*** | 10.58*** |
| | (0.01) | (0.03) | (0.02) | (0.04) | (0.11) | (0.24) | (0.13) | (0.25) |
| Controls | - | yes | - | yes | - | yes | - | yes |
| Observations | 15,426 | 15,426 | 6,411 | 6,411 | 12,188 | 12,188 | 5,755 | 5,755 |
| $R^2$ | 0.06 | 0.17 | 0.05 | 0.20 | 0.00 | 0.17 | 0.02 | 0.20 |
| **Panel C: Advanced Secondary Level** | | | | | | | | |
| Advanced Secondary | 1.50*** | 1.49*** | 1.31*** | 1.36*** | -1.13 | 1.53 | 0.39 | 3.81 |
| | (0.03) | (0.03) | (0.03) | (0.03) | (2.88) | (2.03) | (2.24) | (2.85) |
| Constant | 11.58*** | 10.93*** | 11.93*** | 11.09*** | 11.95*** | 10.93*** | 12.09*** | 10.46*** |
| | (0.01) | (0.03) | (0.01) | (0.04) | (0.15) | (0.30) | (0.17) | (0.30) |
| Controls | - | yes | - | yes | - | yes | - | yes |
| Observations | 16,817 | 16,817 | 7,256 | 7,256 | 13,513 | 13,513 | 6,583 | 6,583 |
| $R^2$ | 0.11 | 0.19 | 0.14 | 0.24 | 0.00 | 0.20 | 0.07 | 0.00 |
| Mean Log Monthly Earnings (TZS) | 11.71 | 11.71 | 12.09 | 12.09 | 11.71 | 11.71 | 12.09 | 12.09 |

*Note:* Data source: Tanzania's 2014 Integrated Labor Force Survey. Sample includes employed males and females of Dar es Salaam aged 15 to 80 years. Monthly earnings are adjusted for inflation with 2012 serving as the base year (Exchange rate in 2012: $1=1,615 TZS). Primary, Ordinary Secondary, and Advanced Secondary refer to people who completed only primary, ordinary secondary, and advanced secondary schooling, respectively. Primary Schooling , Ordinary Secondary and Advanced Secondary Schooling were instrumented with quarterly birth dummy variables (reference group is the last quarter). In columns 2, 4, 6, and 8, we also controlled for gender, experience, and experience square. Standard errors in parentheses. *** Significant at the 1 percent level. ** Significant at the 5 percent level.* Significant at the 10 percent level.

The OLS results clearly show evidence that being a degree holder of the advanced secondary-level is associated with higher monthly earnings by approximately 131 percent to 136 percent in Dar es Salaam. Once we instrument for the schooling level with the quarter of birth variables, the returns to schooling increase. However, some of the coefficients based on the instrumental variable analysis do not pass standard tests of statistical significance. For Dar es Salaam's advanced secondary school level, the OLS coefficient estimate of being an advanced secondary level holder changes from 1.36 to 3.81 (the IV estimate). This estimate change implies that if one obtains an advanced secondary school level, all else equal, one's monthly earnings increase by 381 percent based on the IV estimates as compared to 136 percent based on the OLS estimates.

**Table 7. Returns to Education: First Stage Estimates.**

Panel A: Tanzania mainland

| | Birth Cohort | Mean | Quarter-of-birth effect | | | F-test (p-value) |
|---|---|---|---|---|---|---|
| | | | I | II | III | |
| Years of Education | 1916-1999 | 8.51 | -0.54*** | -0.51*** | -0.08 | 11.03 |
| | | | (0.12) | (0.12) | (0.12) | (0.00) |

Panel B: Dar es Salaam

| | Birth Cohort | Mean | Quarter-of-birth effect | | | F-test (P-value) |
|---|---|---|---|---|---|---|
| | | | I | II | III | |
| Years of Education | 1916-1999 | 10.02 | -0.69*** | -0.53*** | -0.07 | 7.66 |
| | | | (0.17) | (0.17) | (0.18) | (0.00) |

*Note*: Data source: Tanzania's 2014 Integrated Labor Force Survey. The sample includes employed males and females of Dar es Salaam aged 15 to 80 years. Robust standard errors in parentheses. *** Significant at the 1 percent level. ** Significant at the 5 percent level.* Significant at the 10 percent level.

**Table 8. Returns to Education: Instrumental Variable Estimates.**

| Variables | Log of Monthly Earnings (TZS) | | | | | | | |
|---|---|---|---|---|---|---|---|---|
| | OLS estimates | | | | IV estimates | | | |
| | Tanzania mainland | Tanzania mainland | Dar es Salaam | Dar es Salaam | Tanzania mainland | Tanzania mainland | Dar es Salaam | Dar es Salaam |
| | (1) | (2) | (3) | (4) | (5) | (6) | (7) | (8) |
| Years of Schooling | 0.11*** | 0.12*** | 0.10*** | 0.11*** | -0.01 | 0.07* | 0.03 | 0.09** |
| | (0.00) | (0.00) | (0.00) | (0.00) | (0.04) | (0.04) | (0.05) | (0.04) |
| Experience[a] | | 0.05*** | | 0.05*** | | 0.04** | | 0.05** |
| | | (0.00) | | (0.00) | | (0.02) | | (0.02) |
| Experience Squared | | -0.00*** | | -0.00*** | | -0.00*** | | -0.00*** |
| | | (0.00) | | (0.00) | | (0.00) | | (0.00) |
| Gender (1 if male) | | 0.54*** | | 0.57*** | | 0.58*** | | 0.60*** |
| | | (0.02) | | (0.02) | | (0.02) | | (0.03) |
| Constant | 10.82*** | 9.75*** | 11.10*** | 9.94*** | 11.95*** | 10.37*** | 11.84*** | 10.29*** |
| | (0.02) | (0.04) | (0.03) | (0.05) | (0.41) | (0.55) | (0.50) | (0.58) |
| $R^2$ | 0.16 | 0.26 | 0.17 | 0.29 | 0.02 | 0.18 | 0.07 | 0.29 |
| Mean Log Monthly Earnings | 11.71 | 11.71 | 12.09 | 12.09 | 11.71 | 11.71 | 12.09 | 12.09 |
| Observations | 16.817 | 16.817 | 7.256 | 7.256 | 13.513 | 13.513 | 6.583 | 6.583 |

*Note*: Data source: Tanzania's 2014 Integrated Labor Force Survey. Sample includes employed males and females of Dar es Salaam aged 15 to 80 years. Monthly earnings are adjusted for inflation with 2012 serving as the base year (Exchange rate in 2012: $1=1,615 TZS). [a]Experience is calculated by taking the difference of one's age and one's schooling minus six years following Mincer (1974), Boissiere, Knight and Sabot (1985) and Lemieux (2006). Years of schooling was instrumented with quarterly birth dummy variables (reference group is the fourth quarter). In columns 2, 4, 6, and 8, we also controlled for gender, experience, and experience squared. Robust standard errors in parentheses. *** Significant at the 1 percent level. ** Significant at the 5 percent level.* Significant at the 10 percent level.

Using education measured in years of schooling, we find that an additional year of schooling increases earnings from 9 to 11 percent, and that the IV estimates are, in general, slightly lower than the OLS estimates. Measuring schooling as a continuous variable, Table 8 reports the OLS results based on specification (1) in columns 1-4. Columns 5-8 of Table 8 report the IV estimates based on specifications (2)-(3) for both mainland Tanzania and Dar es Salaam with schooling also measured in years of education.[15] The OLS estimates indicate a significant increase in earnings with each additional year of schooling acquired. The point estimate for the mainland indicates that for each additional year of schooling, all else equal, earnings increase by 12 percent and 11 percent respectively for mainland Tanzania and for the Dar es Salaam area. The IV estimates of the returns to schooling for both the mainland and Dar es Salaam are slightly lower than the corresponding OLS estimates. The IV estimate for mainland Tanzania indicates that each additional year of schooling, all else equal, increases earnings by 7 percent. The associated standard errors, however, are large. The IV estimate for Dar es Salaam indicates that each additional year of schooling, all else equal, increases earnings by 9 percent.

---

[15] Table 7 reports the F-test of instruments in the first stage regression and shows that they are jointly statistically significant at the 1-percent level.

Card (2001) summarizes evidence-based studies examining the returns to schooling primarily in developed countries. His summary highlights that IV estimates of the return to schooling typically exceed the corresponding OLS estimates -- often by 20 percent or more (and consistent with the results we report in Table 6). Additionally, one of the few studies that presents credible causal estimates of the returns to schooling in developing countries is Duflo (2001). Exploiting an exogenous variation due to a school construction program in the 1970's, Duflo (2001) examines the return to education in Indonesia.[16] The study reports an OLS estimate of 0.057, an IV estimate without added controls of 0.064, and an IV estimate with controls for district-level enrolment rates of 0.049.

Three explanations can generally account for the discrepancy between the OLS and IV estimates. First, it could be measurement error that accounts for differences in the OLS and IV estimates as measurement error generally generates downward bias in the OLS estimates. Second, it could be that the IV estimates differ from the corresponding OLS estimates because of unobservable differences in characteristics of the "treatment" and "comparison" groups (e.g. "*ability bias*" -- individuals of higher ability tend to get more education causing upwards bias in the OLS estimates). Third, the IV estimation relies on identifying variation based on the so-called group of "compliers", which could have relatively high returns to education (Angrist et al. 1996; Imbens and Rubin 1997). Our results (Table 8 displays the specification consistent with related studies) generally show IV estimates that are slightly lower than the OLS ones. In our case, the discrepancy is likely a product of high positive ability bias in the OLS estimates and because we have missing ILFS data on birth quarters for individuals with high marginal returns to schooling.

---

[16] The program set a target number of primary schools to be built in each of Indonesia's 281 districts, based on the enrollment rate of primary-school age children in the district in 1972. The study shows that average educational attainment rose more quickly in districts that had greater program intensity; measured by the target number of new schools per primary-school age student in the district in 1971. Using a continuous measure of schooling attained, Duflo (2001) reports results based on monthly earnings.

***5.4 Subjective Perceptions and Objective Measures of Returns to Schooling by Education Level***

Next, we examine the average subjective perceptions about earnings using the Dar es Salaam Perceived Returns Survey sample. We follow three main steps to accomplish our main objective in this part of the analysis. First, we compute the mean of the elicited response to the question outlined earlier in section 3.2. This average number will be interesting in its own right and will serve as a comparison to the actual average earnings calculated from the 2014 ILFS survey data. Second, by aggregating the average individual perceptions for the two educational levels – primary and secondary– we can then compute the gap between the elicited subjective perceptions about earnings and the average measured earnings from the ILFS dataset. Finally, based on the Perceived Returns Survey sample, we can examine how various socioeconomic characteristics of survey respondents correlate with the gap between the elicited subjective perceptions about earnings and the average measured earnings from the ILFS dataset.

Table 9 provides data on individual perceptions, elicited by the household head, regarding returns to schooling at each schooling level attained. Table 9 Panel A reports the perceived monthly earnings for males aged 40 to 50 years old who have attained only a primary-level of schooling. Table 9 Panel B reports the perceived monthly earnings for males aged 40 to 50 years old who have attained a secondary-level of schooling. On average, the male workers aged 40 to 50 years old with only primary school education were perceived to earn 81,383 Tanzanian Shillings (≈50 US Dollars) while the average perceived earnings for males aged 40 to 50 years old with only secondary-level education were 172,753 Tanzanian Shillings (≈107 US Dollars).

**Table 9. Perceived Average Monthly Earnings for 40-50 Year-Old Males.**

|  | Monthly Earnings (TZS) |
|---|---|
| **Panel A** |  |
| Primary-level Education Completed Only | 81,382.91 |
|  | (47,711.58) |
| Observations | 1,264 |
| **Panel B** |  |
| Secondary-level Education Completed Only | 172,752.60 |
|  | (74,338.69) |
| Observations | 1,264 |

*Note*: Data source: Cognitive Development Study survey, 2012. Exchange rate in 2012: $1=1,615 TZS. Sample includes household heads of Dar es Salaam who participated in the survey. Dependent variable: Monthly Earnings (TZS) is the household head's perceived average monthly earnings at various education levels for employed males aged 40 to 50 years. Primary-level Education Completed Only and Secondary-level Education Completed Only includes people who passed only primary-level and secondary-level schools, respectively. Standard deviations in parentheses. *** Significant at the 1 percent level. ** Significant at the 5 percent level.* Significant at the 10 percent level.

To be able to compare the subjectively perceived earnings with the actual measured earnings, we obtain the measured returns to education by computing the simple difference in mean earnings by an education level. Although this measure is not likely to be purged of potential econometric concerns, we use this measure rather than estimates adjusted for other covariates or the IV estimates for two reasons. First, estimated returns increase by only 1 percentage points when we account for additional control covariates and the estimates shrink by less than 10 percent when we use the quarter of birth as an instrument. Second, while estimating the returns to schooling precisely is important, our analysis is mainly focused on uncovering factors associated with the gap between actual measured returns and perceived returns based on subjective beliefs. Thus, it is unlikely that using a more precise measure of earnings will influence the actual set of factors that is associated with this gap between the measured and perceived returns.

### 5.5 What Drives the Gap Between Real and Subjective Returns to Schooling?

Using data on the average measured returns and the average perceived returns to schooling, we compute the gap between the two measures and examine what factors drive the wedge between them for men aged 40 to 50 years. Figures 3 and 4 depict the gap in level and in proportional terms.

## Figure 3. Gap between Average Measured and Average Perceived Earnings at Different Levels of Schooling.

**Panel A: by Age of Household Head**

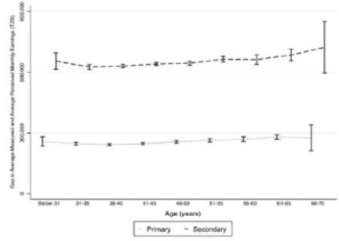

Note: The figure shows how the gap between average measured and average perceived monthly earnings at different school completion levels varies by age of household head. Average measured earnings are the average monthly earnings of 40 to 50 year-old males of Dar es Salaam calculated from Tanzania's 2014 Integrated Labor Force Survey (adjusted for inflation with 2012 serving as the base year). Exchange rate in 2012: $1=1,615 TZS. Average perceived earnings are the average monthly earnings for employed males of Dar es Salaam aged 40 to 50 years, as perceived by the household head surveyed in the 2012 Cognitive Development Study. The variable "Gap" is the difference between the average measured earnings and the average perceived earnings.

**Panel B: by Gender of Household Head**

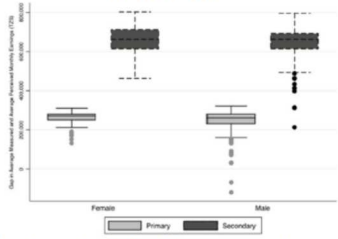

Note: The figure shows how the gap between average measured and average perceived monthly earnings at different school completion levels varies by gender of household head. Average measured earnings are the average monthly earnings of 40 to 50 year-old males of Dar es Salaam calculated from Tanzania's 2014 Integrated Labor Force Survey (adjusted for inflation with 2012 serving as the base year). Exchange rate in 2012: $1=1,615 TZS. Average perceived earnings are the average monthly earnings for employed males of Dar es Salaam aged 40 to 50 years as perceived by the household head surveyed in the 2012 Cognitive Development Study. The variable "Gap" is the difference between the average measured earnings and the average perceived earnings.

**Panel C: by Education of Household Head**

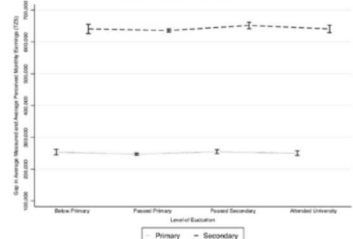

Note: The figure shows how the gap between average measured and average perceived monthly earnings at different school completion levels varies by education level of household head. Average measured earnings are the average monthly earnings of 40 to 50 year-old males of Dar es Salaam calculated from Tanzania's 2014 Integrated Labor Force Survey (adjusted for inflation with 2012 serving in the base year). Exchange rate in 2012: $1=1,615 TZS. Average perceived earnings are the average monthly earnings for employed males of Dar es Salaam aged 40 to 50 years, as perceived by the household head surveyed in the 2012 Cognitive Development Study. The variable "Gap" is the difference between the average measured earnings and the average perceived earnings.

**Panel D: by Household Earnings Deciles**

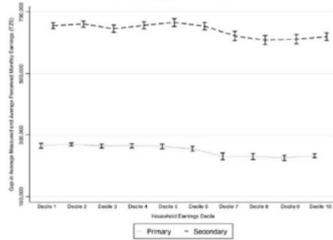

Note: The figure shows how the gap between average measured and average perceived monthly earnings at different school completion levels varies by household earnings deciles: deciles where decile 1 is the poorest group and decile 10 is richest group. Average measured earnings are the average monthly earnings of 40 to 50 year-old males of Dar es Salaam calculated from Tanzania's 2014 Integrated Labor Force Survey (adjusted for inflation with 2012 serving in the base year). Exchange rate in 2012: $1=1,615 TZS. Average perceived earnings are the average monthly earnings for employed males of Dar es Salaam aged 40 to 50 years, as perceived by the household head surveyed in the 2012 Cognitive Development Study. The variable "Gap" is the difference between the average measured earnings and the average perceived earnings.

**Panel E: by Household Asset Value Deciles**

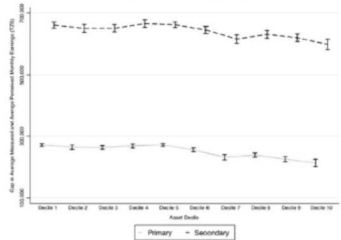

Note: The figure shows how the gap between average measured and average perceived monthly earnings at different school completion levels varies by household asset value deciles (calculated using the asset value) where decile 1 covers the household with lowest asset value and decile 10 covers the household with highest asset value. Average measured earnings are the average monthly earnings of 40 to 50 year-old males of Dar es Salaam calculated from Tanzania's 2014 Integrated Labor Force Survey (adjusted for inflation with 2012 serving as the base year). Exchange rate in 2012: $1=1,615 TZS. Average perceived earnings are the average monthly earnings for employed males of Dar es Salaam aged 40 to 50 years as perceived by the household head surveyed in the 2012 Cognitive Development Study. The variable "Gap" is the difference between the average measured earnings and the average perceived earnings.

**Panel F: by Household Asset Poverty Ranking**

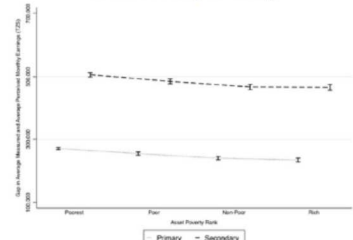

Note: The figure shows how the gap between average measured and average perceived monthly earnings at different school completion levels varies by household asset poverty ranking. We created an asset index score for each household using information on the household condition, value of livestock assets and household asset. Then we divided the household in four groups based on the score where group 1 is the most poorest household and group 4 is the most richest household. Average measured earnings are the average monthly earnings of 40 to 50 year-old males of Dar es Salaam calculated from Tanzania's 2014 Integrated Labor Force Survey (adjusted for inflation with 2012 serving in the base year). Exchange rate in 2012: $1=1,615 TZS. Average perceived earnings are the average monthly earnings for employed males of Dar es Salaam aged 40-50 years as perceived by the household head surveyed in the 2012 Cognitive Development Study. The variable "Gap" is the difference between the average measured earnings and the average perceived earnings.

**Figure 4. Ratio of Average Measured and Average Perceived Earnings at Various Schooling Levels.**

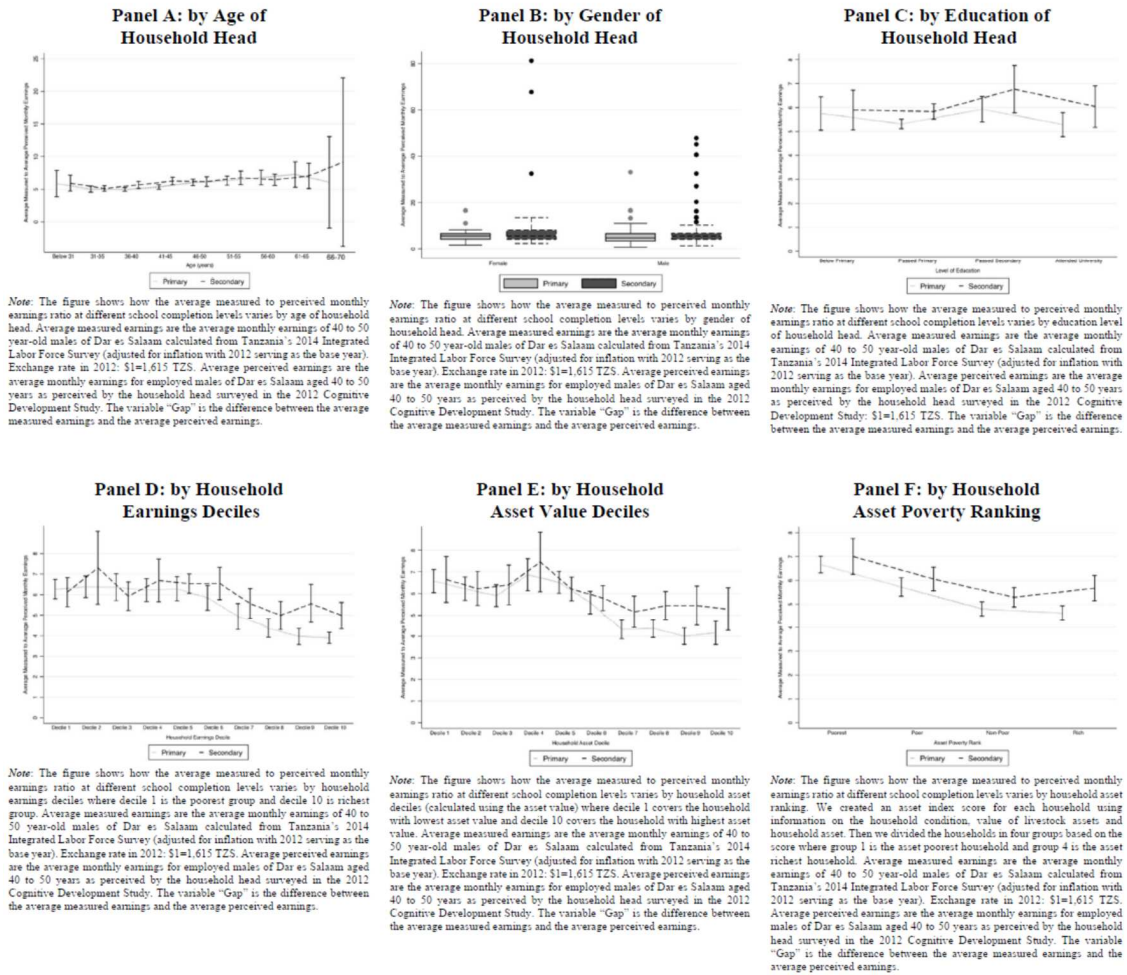

**Panel A: by Age of Household Head**

*Note:* The figure shows how the average measured to perceived monthly earnings ratio at different school completion levels varies by age of household head. Average measured earnings are the average monthly earnings of 40 to 50 year-old males of Dar es Salaam calculated from Tanzania's 2014 Integrated Labor Force Survey (adjusted for inflation with 2012 serving as the base year). Exchange rate in 2012: $1≈1,615 TZS. Average perceived earnings are the average monthly earnings for employed males of Dar es Salaam aged 40 to 50 years as perceived by the household head surveyed in the 2012 Cognitive Development Study. The variable "Gap" is the difference between the average measured earnings and the average perceived earnings.

**Panel B: by Gender of Household Head**

*Note:* The figure shows how the average measured to perceived monthly earnings ratio at different school completion levels varies by gender of household head. Average measured earnings are the average monthly earnings of 40 to 50 year-old males of Dar es Salaam calculated from Tanzania's 2014 Integrated Labor Force Survey (adjusted for inflation with 2012 serving as the base year). Exchange rate in 2012: $1≈1,615 TZS. Average perceived earnings are the average monthly earnings for employed males of Dar es Salaam aged 40 to 50 years as perceived by the household head surveyed in the 2012 Cognitive Development Study. The variable "Gap" is the difference between the average measured earnings and the average perceived earnings.

**Panel C: by Education of Household Head**

*Note:* The figure shows how the average measured to perceived monthly earnings ratio at different school completion levels varies by education level of household head. Average measured earnings are the average monthly earnings of 40 to 50 year-old males of Dar es Salaam calculated from Tanzania's 2014 Integrated Labor Force Survey (adjusted for inflation with 2012 serving in the base year). Average perceived earnings are the average monthly earnings for employed males of Dar es Salaam aged 40 to 50 years as perceived by the household head surveyed in the 2012 Cognitive Development Study. The variable "Gap" is the difference between the average measured earnings and the average perceived earnings.

**Panel D: by Household Earnings Deciles**

*Note:* The figure shows how the average measured to perceived monthly earnings ratio at different school completion levels varies by household earnings decile; where decile 1 is the poorest group and decile 10 is richest group. Average measured earnings are the average monthly earnings of 40 to 50 year-old males of Dar es Salaam calculated from Tanzania's 2014 Integrated Labor Force Survey (adjusted for inflation with 2012 serving as the base year). Exchange rate in 2012: $1≈1,615 TZS. Average perceived earnings are the average monthly earnings for employed males of Dar es Salaam aged 40 to 50 years as perceived by the household head surveyed in the 2012 Cognitive Development Study. The variable "Gap" is the difference between the average measured earnings and the average perceived earnings.

**Panel E: by Household Asset Value Deciles**

*Note:* The figure shows how the average measured to perceived monthly earnings ratio at different school completion levels varies by household asset deciles (calculated using the asset value) where decile 1 covers the household with lowest asset value and decile 10 covers the household with highest asset value. Average measured earnings are the average monthly earnings of 40 to 50 year-old males of Dar es Salaam calculated from Tanzania's 2014 Integrated Labor Force Survey (adjusted for inflation with 2012 serving in the base year). Exchange rate in 2012: $1≈1,615 TZS. Average perceived earnings are the average monthly earnings for employed males of Dar es Salaam aged 40 to 50 years as perceived by the household head surveyed in the 2012 Cognitive Development Study. The variable "Gap" is the difference between the average measured earnings and the average perceived earnings.

**Panel F: by Household Asset Poverty Ranking**

*Note:* The figure shows how the average measured to perceived monthly earnings ratio at different school completion levels varies by household asset ranking. We created an asset index score for each household using information on the household condition, value of livestock assets and household asset. Then we divided the households in four groups based on the score where group 1 is the most poorest household and group 4 is the most richest household. Average measured earnings are the average monthly earnings of 40 to 50 year-old males of Dar es Salaam calculated from Tanzania's 2014 Integrated Labor Force Survey (adjusted for inflation with 2012 serving in the base year). Exchange rate in 2012: $1≈1,615 TZS. Average perceived earnings are the average monthly earnings for employed males of Dar es Salaam aged 40 to 50 years as perceived by the household head surveyed in the 2012 Cognitive Development Study. The variable "Gap" is the difference between the average measured earnings and the average perceived earnings.

On average, individuals in Dar es Salaam expected monthly earnings of 81,383 Tanzanian Shillings (≈50 US Dollars) for individuals aged 40 to 50 years old who only completed primary school. In contrast, the average measured earnings for individuals 40 to 50 years old with primary-level schooling was 309,332 Tanzanian Shillings (≈191 US Dollars) in the ILFS earnings data for the Dar es Salaam area. There was a similar pattern between perceived and measured earnings for individuals with secondary schooling: on average, individuals expected monthly earnings of 172,753 Tanzanian Shillings (≈107 US Dollars) for individuals who have completed only secondary-level schooling. The average measured earnings for individuals with only secondary-level schooling in Dar es Salaam, however, was 834,350 Tanzanian Shillings (≈516 US Dollars). In summary, comparing the perceived earnings with the average measured earnings, individuals seem to underestimate the average earnings for individuals with primary-level schooling by about 227,949 TZS (or 74 percent) and underestimate the average earnings for individuals with secondary-level schooling only by about 661,597 TZS (or 79 percent).

We specifically examine how individual factors of the respondent, such as age, gender, educational level, income decile, asset decile and asset poverty ranking relate to the discrepancy between measured returns and subjectively perceived returns to schooling. Figures 3 and 4 depict several patterns on how each of the socioeconomic factors relate to the magnitude of the discrepancy between measured and perceived earnings. Figure 3 depicts level changes whereas Figure 4 depicts proportional changes. In general, the slope of each figure in the panels represents how responsive the difference between the average measured earnings and subjectively perceived earnings to schooling is to unit changes in the variable on the x-axis. Figure 4, in particular, highlights three stylized patterns. First, Figure 4, Panel C shows that a change of educational attainment from primary-level to secondary-level contributes to a large increase in the gap between measured and perceived returns. Second, Figure 4 Panel A shows an increase in the gap as the respondent's age increases. Third, Panels D - F display how one's income decile, one's asset value and one's asset poverty relate to changes in the gap between the measured average earnings and the subjectively elicited beliefs about average earnings. Panels D and F show that the gap is the largest among highest the poverty levels.

We formalize the graphical analyses presented in Figure 3 and Figure 4 with the use of ordinary least squares analysis. We analyse the gap between average measured earnings and the earnings based on subjective beliefs as a dependent variable and regress that gap outcome on the set of individual socioeconomic covariates. Table 10 reports results from OLS specifications.

In terms of statistical significance, the three most powerful predictors of the gap between measured returns and subjectively perceived returns to schooling are: respondent's age, whether one has a secondary school or university-level education, and one's poverty status (*i.e.*, asset poverty and earnings).

**Table 10. Determinants of the Gap Between the Measured and Perceived Average Monthly Earnings for 40-50 Year-Old Males in Dar es Salaam.**

| | Gap between the Measured and Perceived Average Monthly Earnings (TZS) | |
| --- | --- | --- |
| | Primary School Completed (1) | Secondary School Completed (2) |
| Household Head Age (years) | 936.39*** | 1,576.42*** |
| | (181.95) | (333.45) |
| Household Head Gender (1 if male) | -5,939.17 | -23,509.14** |
| | (6,441.03) | (10,714.64) |
| Household Head Marital Status (1 if married) | -4,190.86 | 9,412.37 |
| | (6,225.74) | (10,130.29) |
| Head Completed Primary Schooling (binary) | -2,379.49 | 1,588.40 |
| | (4,251.31) | (7,948.72) |
| Head Completed Secondary Schooling (binary) | 10,601.18** | 22,274.25** |
| | (4,894.80) | (9,078.45) |
| Head Attended University (binary) | 9,209.36* | 14,566.14 |
| | (5,346.77) | (9,768.33) |
| Education of Spouse (years of schooling) | -120.75* | -0.69 |
| | (72.72) | (132.94) |
| Asset Poverty Rank (indicator for group 2)[a] | -13,527.61*** | -17,398.85*** |
| | (3,497.21) | (5,830.09) |
| Asset Poverty Rank (indicator for group 3)[a] | -28,175.85*** | -36,133.17*** |
| | (3,422.64) | (5,830.37) |
| Asset Poverty Rank (indicator for group 4)[a] | -35,771.15*** | -39,966.19*** |
| | (4,079.90) | (6,662.37) |
| Household Monthly Earnings (log) | -1,492.09*** | -1,594.12*** |
| | (576.28) | (565.87) |
| Constant | 246,734.53*** | 608,739.67*** |
| | (11,147.80) | (18,275.92) |
| $R^2$ | 0.13 | 0.09 |
| Observations | 1,210 | 1,211 |

*Note:* Data source: Cognitive Development Study survey, 2012. Exchange rate in 2012: $1=1,615 TZS. Sample includes household heads of Dar es Salaam participated in the survey. [a]We created an asset score for each household based on the household condition, the value of livestock asset, and household assets. Households were then divided into four groups based on the relative score (Group 1 is the asset poorest household and Group 4 is the asset richest household). *** Significant at the 1 percent level. ** Significant at the 5 percent level.* Significant at the 10 percent level.

We also examine the coefficient magnitudes for factors associated with the gap between measured returns and subjectively perceived returns to schooling. The magnitudes of these coefficients are particularly interesting given that they can shed light on potential policy interventions focused on correcting individual misperceptions. Interestingly, the largest coefficients are associated with the following variables: individual earnings and educational attainment. The lowest earners as well as individuals in the lowest decile based on asset poverty are the groups that underestimate educational returns the most. Interestingly, secondary school degree holders (relative to primary school level holders or no degree holders) also underestimate educational returns.

## 6. Discussion and Conclusion

Building on a very extensive literature in human capital, Manski (1993) posited that individual beliefs regarding educational returns could be a powerful determinant of individual demand for more schooling. In this paper, we use individual-level data from two surveys in Tanzania to estimate the returns to primary and secondary schooling and to examine whether subjective perceptions regarding the monetary returns to schooling differ from measured average returns for these two schooling levels. We also examine what factors are associated with the gap between actual measures of educational returns and subjective perceptions.

We find that each additional year of schooling in Tanzania, all else equal, increases earnings by 11 percent in the OLS estimates and by 9 percent in the IV estimates. Using data from Dar es Salaam's Perceived Returns Survey, we also examine the individual subjective perceptions regarding the average earnings associated with two levels of schooling. We find that survey respondents underestimate the average earnings for workers with primary-level schooling by 74 percent and that survey respondents underestimate the average earnings for individuals with secondary-level schooling by approximately 79 percent. Using limited data on the socioeconomic characteristics of the survey respondent, we then examine the role that each individual factor plays in driving the discrepancy between the measured average earnings and the subjectively perceived average earnings.

We find three powerful predictors that drive the gap between the subjectively perceived average earnings and the actual average measured earnings: the respondent's age, whether one has a secondary school or university-level education and one's poverty status (based asset poverty and earnings). Perhaps most policy-relevant is the fact that the largest effects, in terms of the estimated coefficient magnitudes, driving the discrepancy between measured earnings and subjective beliefs about earnings are associated with one's own earnings and one's own educational attainment. The lowest earners as well as the lowest decile in asset poverty are the two groups of individuals who underestimate the average earnings the most. Surprisingly, secondary school degree holders (relative to primary school degree holders or no degree holders) also underestimate educational returns. Although our measure of the returns may still be biased, the individuals' implied estimates of the returns are so low -- about 3-4 percent per year of secondary schooling -- that unless we believe our estimates of the actual educational returns are highly biased such that the true returns in Tanzania are dramatically lower than the returns we estimate in this paper, it seems likely that individuals do in fact underestimate the true returns to schooling.

Finally, we note that within the Becker human capital framework, there are numerous reasons – other than low perceived returns to schooling – that may drive the equilibrium in which individuals receive low levels of education. Such factors, for example poverty and credit constraints, have long been considered significant impediments to schooling, especially in Sub-Saharan countries. However, relaxing these other constraints is unlikely to be a particularly cost-effective strategy. The results of this paper point to an alternative cost-effective policy approach, in which a policy targets groups that underestimate the measured educational returns the most. A targeted low-cost informational intervention among the lowest earners and the lowest asset poor decile is likely to result in a powerful impact on these groups' demand for more schooling.

# Appendix A: Supplementary Tables

## Table A1. Mean Educational Attainment (by Gender).

| | Tanzania Mainland | | | Dar es Salaam | | |
|---|---|---|---|---|---|---|
| | Male | Female | Difference | Male | Female | Difference |
| | (1) | (2) | (3) | (4) | (5) | (6) |
| Never Attended School (1 if schooling is zero) | 0.06 | 0.14 | -0.07*** | 0.02 | 0.06 | -0.04*** |
| | (0.24) | (0.34) | (0.00) | (0.15) | (0.23) | (0.00) |
| Incomplete Primary Education (1 if schooling <7) | 0.61 | 0.38 | 0.22*** | 0.68 | 0.48 | 0.19** |
| | (0.48) | (0.48) | (0.01) | (0.46) | (0.50) | (0.03) |
| Attained Primary Education Only (1 if schooling>=7) | 0.78 | 0.72 | 0.06* | 0.89 | 0.84 | 0.05*** |
| | (0.42) | (0.44) | (0.01) | (0.31) | (0.36) | (0.01) |
| Attained Secondary Ordinary Level Education (1 if schooling>=11) | 0.20 | 0.16 | 0.04*** | 0.29 | 0.23 | 0.06*** |
| | (0.39) | (0.36) | (0.00) | (0.45) | (0.42) | (0.01) |
| Attained Secondary Advanced Level Education (1 if schooling>=13) | 0.09 | 0.06 | 0.04*** | 0.14 | 0.09 | 0.05*** |
| | (0.29) | (0.23) | (0.00) | (0.35) | (0.28) | (0.01) |
| Years of Schooling | 9.15 | 7.95 | 1.20*** | 10.68 | 9.45 | 1.23*** |
| | (5.00) | (5.00) | (0.06) | (4.93) | (4.87) | (0.09) |
| Observations | 13,746 | 15,289 | 29,035 | 5,809 | 6,501 | 12,310 |

*Note:* Data source: Tanzania's 2014 Integrated Labor Force Survey. The sample includes males and females aged 15 years or more. Primary, Ordinary Secondary and Advanced Secondary refers to people who completed only Primary, Ordinary Secondary or Advanced Secondary schooling, respectively. Years of schooling denotes the total years of schooling. Standard errors in parentheses. *** Significant at the 1 percent level. ** Significant at the 5 percent level.* Significant at the 10 percent level.

**Table A2. Economic Activities and Working Hours**
**(Distribution by Gender and Education Level).**

| Variables | Economic Activities in Last Twelve Months[a] (dummy) | | | | | | Monthly Working Hours[b] (in Economic Activities) | | | | | |
|---|---|---|---|---|---|---|---|---|---|---|---|---|
| | Tanzania Mainland | | | Dar es Salaam | | | Tanzania Mainland | | | Dar es Salaam | | |
| | Male | Female | Diff. | Male | Female | Diff. | Male | Female | Diff. | Male | Female | Diff. |
| | (1) | (2) | (3) | (4) | (5) | (6) | (7) | (8) | (9) | (10) | (11) | (12) |
| Never Attended School (1 if schooling is zero) | 0.84 | 0.71 | 0.13*** | 0.65 | 0.42 | 0.24*** | 167.75 | 138.63 | 29.11*** | 235.71 | 181.53 | 54.17*** |
| | (0.37) | (0.45) | (0.02) | (0.47) | (0.49) | (0.05) | (97.62) | (90.77) | (4.32) | (86.46) | (94.00) | (12.10) |
| Incomplete Primary Education (1 if schooling <7) | 0.85 | 0.73 | 0.12*** | 0.71 | 0.48 | 0.24*** | 178.58 | 144.60 | 33.97*** | 234.25 | 192.66 | 41.59*** |
| | (0.36) | (0.44) | (0.02) | (0.45) | (0.50) | (0.04) | (107.86) | (98.22) | (4.52) | (113.4) | (111.53) | (11.84) |
| Attained Primary Education Only (1 if schooling>=7) | 0.84 | 0.67 | 0.17*** | 0.78 | 0.50 | 0.28*** | 215.33 | 175.73 | 39.59*** | 253.46 | 209.76 | 43.70*** |
| | (0.36) | (0.45) | (0.01) | (0.41) | (0.49) | (0.01) | (105.52) | (100.96) | (1.84) | (95.60) | (102.78) | (3.02) |
| Attained Secondary Ordinary Level Education (1 if schooling>=11) | 0.70 | 0.55 | 0.15*** | 0.65 | 0.46 | 0.19*** | 224.50 | 188.09 | 36.41*** | 239.28 | 204.07 | 35.21*** |
| | (0.45) | (0.49) | (0.01) | (0.47) | (0.49) | (0.02) | (97.34) | (95.91) | (3.58) | (92.10) | (95.76) | (4.87) |
| Attained Secondary Advanced Level Education (1 if schooling>=13) | 0.72 | 0.63 | 0.11*** | 0.68 | 0.56 | 0.12*** | 197.56 | 175.95 | 21.60*** | 195.30 | 171.61 | 23.69*** |
| | (0.44) | (0.48) | (0.02) | (0.47) | (0.49) | (0.03) | (75.64) | (66.87) | (3.84) | (73.08) | (62.05) | (4.70) |
| All | 0.81 | 0.66 | 0.14*** | 0.72 | 0.49 | 0.23*** | 208.42 | 168.95 | 39.46*** | 241.58 | 202.68 | 38.89*** |
| | (0.39) | (0.47) | (0.01) | (0.44) | (0.49) | (0.49) | (103.20) | (98.49) | (1.38) | (94.85) | (98.97) | (2.27) |
| Observations | 13,746 | 15,289 | 29,035 | 5,809 | 6,501 | 12,310 | 11,089 | 10,118 | 21,207 | 4,231 | 3,208 | 7,439 |

*Note:* Tanzania's 2014 Integrated Labor Force Survey. The sample includes males and females aged 15 years or more. [a]Economic activities in last twelve months denotes whether or not the person was involved in any work or activities for pay, profit, barter or home use in last twelve months. [b]Monthly working hours denotes the hours a person spent per month in economic activities. Mean Monthly Working Hours is calculated for people who were economically active in last 12 months. *** Significant at the 1 percent level. ** Significant at the 5 percent level.* Significant at the 10 percent level.

**Table A3. Returns to Education: By Gender and Working Hours.**

| | Log of Monthly Earnings (TZS) | | | |
| --- | --- | --- | --- | --- |
| | Tanzania mainland | | Dar es Salaam | |
| | (1) | (2) | (3) | (4) |
| Experience[a] | 0.05*** | 0.05*** | 0.05*** | 0.06*** |
| | (0.00) | (0.00) | (0.00) | (0.00) |
| Experience Squared | -0.00*** | -0.00*** | -0.00*** | -0.00*** |
| | (0.00) | (0.00) | (0.00) | (0.00) |
| Gender (1 if male) | 0.59*** | 0.52*** | 0.62*** | 0.55*** |
| | (0.04) | (0.04) | (0.09) | (0.09) |
| Primary (1 if schooling>=7) | 0.41*** | 0.35*** | 0.27*** | 0.25*** |
| | (0.04) | (0.04) | (0.07) | (0.07) |
| Ordinary Secondary (1 if schooling>=11) | 1.36*** | 1.31*** | 1.07*** | 1.08*** |
| | (0.05) | (0.05) | (0.08) | (0.07) |
| Advanced Secondary (1 if schooling>=13) | 2.38*** | 2.38*** | 2.16*** | 2.22*** |
| | (0.05) | (0.05) | (0.08) | (0.08) |
| Gender Dummy * Primary (1 if schooling>=7) | 0.05 | 0.04 | 0.07 | 0.07 |
| | (0.05) | (0.05) | (0.09) | (0.09) |
| Gender Dummy * Ordinary Secondary (1 if schooling>=11) | -0.17*** | -0.19*** | -0.15 | -0.14 |
| | (0.06) | (0.06) | (0.10) | (0.10) |
| Gender Dummy * Advanced Secondary (1 if schooling>=13) | -0.44*** | -0.42*** | -0.45*** | -0.42*** |
| | (0.07) | (0.07) | (0.11) | (0.11) |
| Monthly Working Hours[b] | | 0.00*** | | 0.00*** |
| | | (0.00) | | (0.00) |
| Constant | 10.15*** | 9.73*** | 10.46*** | 10.10*** |
| | (0.04) | (0.04) | (0.07) | (0.08) |
| $R^2$ | 0.27 | 0.3 | 0.3 | 0.32 |
| Mean Log Monthly Earnings (TZS) | 11.71 | 11.71 | 11.71 | 11.71 |
| Observations | 16,817 | 16,817 | 7,256 | 7,256 |

*Note*: Tanzania's 2014 Integrated Labor Force Survey. Monthly earnings are adjusted for inflation with 2012 serving as the base year (Exchange rate in 2012: $1=1,615 TZS). [a]Experience is calculated by taking the difference of one's age and one's schooling minus six years [this is the standard approach in the literature following Mincer (1974), Boissiere, Knight and Sabot (1985) and Lemieux (2006)]. Primary, Ordinary Secondary and Advanced Secondary refers to people who completed only Primary, Ordinary Secondary or Advanced Secondary schooling, respectively. Standard errors in parentheses. [b]Monthly working hours denotes the hours a person spent per month in economic activities. *** Significant at the 1 percent level. ** Significant at the 5 percent level.* Significant at the 10 percent level.

**Table A4. Returns to Education: By Gender and Working Hours.**

| | Log of Monthly Earnings (TZS) | | | | | |
| | Tanzania mainland | | | Dar es Salaam | | |
| | (1) | (2) | (3) | (4) | (5) | (6) |
|---|---|---|---|---|---|---|
| Experience[a] | 0.047*** | 0.048*** | 0.048*** | 0.054*** | 0.056*** | 0.057*** |
| | [0.002] | [0.002] | [0.002] | [0.003] | [0.003] | [0.003] |
| Experience Squared | -0.001*** | -0.001*** | -0.001*** | -0.001*** | -0.001*** | -0.001*** |
| | [0.000] | [0.000] | [0.000] | [0.000] | [0.000] | [0.000] |
| Gender (1 if male) | 0.559*** | 0.480*** | 0.508*** | 0.588*** | 0.520*** | 0.654*** |
| | [0.017] | [0.017] | [0.031] | [0.024] | [0.024] | [0.054] |
| Primary (1 if schooling>=7) | 0.441*** | 0.250*** | 0.251*** | 0.315*** | 0.207* | 0.182 |
| | [0.026] | [0.050] | [0.050] | [0.048] | [0.117] | [0.118] |
| Ordinary Secondary (1 if schooling>=11) | 1.263*** | 1.355*** | 1.333*** | 0.986*** | 1.104*** | 1.053*** |
| | [0.033] | [0.073] | [0.073] | [0.053] | [0.130] | [0.131] |
| Advanced Secondary (1 if schooling>=13) | 2.107*** | 2.314*** | 2.234*** | 1.879*** | 2.018*** | 1.877*** |
| | [0.037] | [0.102] | [0.102] | [0.057] | [0.161] | [0.162] |
| Monthly Working Hours[b] | | 0.002*** | 0.002*** | | 0.001*** | 0.001*** |
| | | [0.000] | [0.000] | | [0.000] | [0.000] |
| Monthly Working Hours*Primary | | 0.001*** | 0.001** | | 0.000 | 0.001 |
| | | [0.000] | [0.000] | | [0.000] | [0.001] |
| Monthly Working Hours*Ordinary Secondary | | -0.001** | -0.000 | | -0.000 | 0.000 |
| | | [0.000] | [0.000] | | [0.001] | [0.001] |
| Monthly Working Hours*Advanced Secondary | | -0.001** | 0.001 | | -0.000 | 0.002** |
| | | [0.000] | [0.001] | | [0.001] | [0.001] |
| Monthly Working Hours*Primary*Gender Dummy | | | 0.000 | | | -0.000 |
| | | | [0.000] | | | [0.000] |
| Monthly Working Hours*Ordinary Secondary*Gender Dummy | | | -0.001*** | | | -0.001*** |
| | | | [0.000] | | | [0.000] |
| Monthly Working Hours*Advanced Secondary*Gender Dummy | | | -0.002*** | | | -0.003*** |
| | | | [0.000] | | | [0.000] |
| Constant | 10.168*** | 9.796*** | 9.782*** | 10.482*** | 10.146*** | 10.104*** |
| | [0.035] | [0.049] | [0.051] | [0.056] | [0.115] | [0.117] |
| $R^2$ | 0.267 | 0.303 | 0.305 | 0.298 | 0.316 | 0.32 |
| Mean Log Monthly Earnings (TZS) | 11.71 | 11.71 | 11.71 | 11.71 | 11.71 | 11.71 |
| Observations | 16,817 | 16,817 | 16,817 | 7,256 | 7,256 | 7,256 |

*Note*: Tanzania's 2014 Integrated Labor Force Survey. Monthly earnings are adjusted for inflation with 2012 serving as the base year (Exchange rate in 2012: $1=1,615 TZS). [a]Experience is calculated by taking the difference of one's age and one's schooling minus six years [this is the standard approach in the literature following Mincer (1974), Boissiere, Knight and Sabot (1985) and Lemieux (2006)]. [b]Monthly working hours denotes the hours a person spent per month in economic activities. Primary, Ordinary Secondary and Advanced Secondary refers to people who completed only Primary, Ordinary Secondary or Advanced Secondary schooling, respectively. Standard errors in parentheses. *** Significant at the 1 percent level. ** Significant at the 5 percent level.* Significant at the 10 percent level.

**Table A5. Sectors of Employment[a] (Distribution by Gender and Education Level).**

| Sectors of Employment | Proportion of Economic Activities[b] | Proportion of Males in Sector | Proportion of total Males | Proportion of total Females | Primary Education Completed Proportion | Ordinary Secondary Completed Proportion | Advanced Secondary Completed Proportion | Mean Monthly Earnings (in TZS) |
|---|---|---|---|---|---|---|---|---|
| | (1) | (2) | (3) | (4) | (5) | (6) | (7) | (8) |
| Government Worker | 0.06 | 0.61 | 0.07 | 0.05 | 0.92 | 0.68 | 0.54 | 596,797.60 |
| Agriculture | 0.42 | 0.48 | 0.38 | 0.45 | 0.62 | 0.06 | 0.01 | 77,109.81 |
| Private Sector Employed | 0.20 | 0.63 | 0.24 | 0.15 | 0.89 | 0.25 | 0.11 | 253,290.40 |
| Private Informal Sector | 0.30 | 0.50 | 0.29 | 0.31 | 0.85 | 0.17 | 0.11 | 267,803.80 |
| Household Duties | 0.03 | 0.36 | 0.02 | 0.04 | 0.85 | 0.21 | 0.03 | 56,784.11 |
| Observations | 21,207 | 21,207 | 11,089 | 10,118 | 21,207 | 21,207 | 21,207 | 21,207 |

*Note*: Tanzania's 2014 Integrated Labor Force Survey. The sample includes males and females aged 15 years or more. [a]Classification of employment by sector is based on the Analytical Report of Tanzania's 2014 Integrated Labor Force Survey. The sectors are government worker (central and local government, parastatal organizations etc.), agriculture, private informal sector, private other sector, and household duties. [b]Economic activities denote a person's involvement in any work or activities for pay, profit, barter or home use in last twelve months. Primary, Ordinary Secondary and Advanced Secondary refers to people who completed only Primary, Ordinary Secondary or Advanced Secondary schooling, respectively. *** Significant at the 1 percent level. ** Significant at the 5 percent level.* Significant at the 10 percent level.

**Table A6. Returns to Education: By Sector of Employment[a].**

| | Log of Monthly Earnings (TZS) | | | |
|---|---|---|---|---|
| | Tanzania mainland | | Dar es Salaam | |
| | (1) | (2) | (3) | (4) |
| Experience[b] | 0.05*** | 0.05*** | 0.05*** | 0.05*** |
| | (0.00) | (0.00) | (0.00) | (0.00) |
| Experience Squared | -0.00*** | -0.00*** | -0.00*** | -0.00*** |
| | (0.00) | (0.00) | (0.00) | (0.00) |
| Gender (1 if male) | 0.59*** | 0.65*** | 0.62*** | 0.61*** |
| | (0.04) | (0.04) | (0.09) | (0.09) |
| Primary (1 if schooling>=7) | 0.41*** | 0.27*** | 0.27*** | 0.25*** |
| | (0.04) | (0.04) | (0.07) | (0.07) |
| Ordinary Secondary (1 if schooling>=11) | 1.36*** | 1.06*** | 1.07*** | 1.01*** |
| | (0.05) | (0.05) | (0.08) | (0.08) |
| Advanced Secondary (1 if schooling>=13) | 2.38*** | 1.88*** | 2.16*** | 1.99*** |
| | (0.05) | (0.06) | (0.08) | (0.09) |
| Gender Dummy * Primary (1 if schooling>=7) | 0.05 | 0.02 | 0.07 | 0.07 |
| | (0.05) | (0.05) | (0.09) | (0.09) |
| Gender Dummy * Ordinary Secondary (1 if schooling>=11) | -0.17*** | -0.23*** | -0.15 | -0.14 |
| | (0.06) | (0.06) | (0.10) | (0.10) |
| Gender Dummy * Advanced Secondary (1 if schooling>=13) | -0.44*** | -0.46*** | -0.45*** | -0.41*** |
| | (0.07) | (0.06) | (0.11) | (0.11) |
| Government Worker | | 0.85*** | | 0.66*** |
| | | (0.09) | | (0.15) |
| Agriculture | | -0.27*** | | 0.20 |
| | | (0.08) | | (0.16) |
| Private Sector Employed | | 0.44*** | | 0.38*** |
| | | (0.08) | | (0.14) |
| Private Informal Sector | | 0.36*** | | 0.34** |
| | | (0.08) | | (0.15) |
| Constant | 10.15*** | 10.03*** | 10.46*** | 10.14*** |
| | (0.04) | (0.09) | (0.07) | (0.16) |
| $R^2$ | 0.27 | 0.33 | 0.30 | 0.31 |
| Mean Log Monthly Earnings (TZS) | 11.71 | 11.71 | 11.71 | 11.71 |
| Observations | 16,817 | 16,817 | 7,256 | 7,256 |

*Note*: Tanzania's 2014 Integrated Labor Force Survey. The sample includes males and females aged 15 years or more. [a]Classification of employment by sectors is based on the Analytical Report of Tanzania's 2014 Integrated Labor Force Survey. The sectors are government worker (central and local government, parastatal organizations etc.), agriculture, private informal sector, private other sector, and household duties. The reference category is 'Household Duites'. [b]Experience is calculated by taking the difference of one's age and one's schooling minus six years [this is the standard approach in the literature following Mincer (1974), Boissiere, Knight and Sabot (1985) and Lemieux (2006)]. Primary, Ordinary Secondary and Advanced Secondary refers to people who completed only Primary, Ordinary Secondary or Advanced Secondary schooling, respectively. Standard errors in parentheses. *** Significant at the 1 percent level. ** Significant at the 5 percent level.* Significant at the 10 percent level.

**Table A7. Returns to Education: By Sector of Employment[a].**

| | Log of Monthly Earnings (TZS) | | | | | |
| | Tanzania mainland | | | Dar es Salaam | | |
| | (1) | (2) | (3) | (4) | (5) | (6) |
|---|---|---|---|---|---|---|
| Experience[b] | 0.05*** | 0.05*** | 0.05*** | 0.05*** | 0.06*** | 0.06*** |
| | (0.00) | (0.00) | (0.00) | (0.00) | (0.00) | (0.00) |
| Experience Squared | -0.00*** | -0.00*** | -0.00*** | -0.00*** | -0.00*** | -0.00*** |
| | (0.00) | (0.00) | (0.00) | (0.00) | (0.00) | (0.00) |
| Gender (1 if male) | 0.54*** | 0.58*** | 0.71*** | 0.57*** | 0.58*** | 0.73*** |
| | (0.02) | (0.02) | (0.04) | (0.02) | (0.02) | (0.06) |
| Years of Schooling | 0.12*** | 0.09*** | 0.09*** | 0.11*** | 0.08* | 0.08* |
| | (0.00) | (0.02) | (0.02) | (0.00) | (0.05) | (0.05) |
| Government Worker | | 0.81*** | 0.77*** | | 0.68* | 0.6 |
| | | (0.20) | (0.20) | | (0.36) | (0.36) |
| Agriculture | | 0.10 | 0.06 | | 0.34 | 0.34 |
| | | (0.19) | (0.19) | | (0.36) | (0.36) |
| Private Sector Employed | | 0.20 | 0.16 | | 0.03 | -0.04 |
| | | (0.19) | (0.19) | | (0.34) | (0.34) |
| Private Informal Sector | | 0.38** | 0.36* | | 0.27 | 0.23 |
| | | (0.19) | (0.19) | | (0.34) | (0.34) |
| Government Worker*Years of Schooling | | 0.01 | 0.03 | | 0.01 | 0.04 |
| | | (0.02) | (0.02) | | (0.05) | (0.05) |
| Agriculture*Years of Schooling | | -0.05** | -0.04* | | -0.02 | 0.02 |
| | | (0.02) | (0.02) | | (0.05) | (0.05) |
| Private Sector Employed*Years of Schooling | | 0.02 | 0.04* | | 0.04 | 0.05 |
| | | (0.02) | (0.02) | | (0.05) | (0.05) |
| Private Informal Sector*Years of Schooling | | -0.01 | 0.00 | | 0.01 | 0.01 |
| | | (0.02) | (0.02) | | (0.05) | (0.05) |
| Gender*Government Worker*Years of Schooling | | | -0.04*** | | | -0.03*** |
| | | | (0.00) | | | (0.00) |
| Gender*Agriculture*Years of Schooling | | | -0.03*** | | | -0.07*** |
| | | | (0.01) | | | (0.02) |
| Gender*Private Sector Employed*Years of Schooling | | | -0.02*** | | | -0.02*** |
| | | | (0.00) | | | (0.01) |
| Gender*Private Informal Sector*Years of Schooling | | | 0.00 | | | 0.00 |
| | | | (0.00) | | | (0.01) |
| Constant | 9.75*** | 9.65*** | 9.61*** | 9.94*** | 9.84*** | 9.81*** |
| | (0.03) | (0.19) | (0.19) | (0.05) | (0.34) | (0.34) |
| $R^2$ | 0.26 | 0.33 | 0.33 | 0.29 | 0.31 | 0.31 |
| Mean Log Monthly Earnings (TZS) | 11.71 | 11.71 | 11.71 | 11.71 | 11.71 | 11.71 |
| Observations | 16,817 | 16,817 | 16,817 | 7,256 | 7,256 | 7,256 |

*Note*: Tanzania's 2014 Integrated Labor Force Survey. The sample includes males and females aged 15 years or more. [a]Classification of employment by sectors is based on the Analytical Report of Tanzania's 2014 Integrated Labor Force Survey. The sectors are government worker (central and local government, parastatal organizations etc.), agriculture, private informal sector, private other sector, and household duties. The reference category is 'Household Duites'. [b]Experience is calculated by taking the difference of one's age and one's schooling minus six years [this is the standard approach in the literature following Mincer (1974), Boissiere, Knight and Sabot (1985) and Lemieux (2006)]. Standard errors in parentheses. *** Significant at the 1 percent level. ** Significant at the 5 percent level.* Significant at the 10 percent level.

**Table A8. Occupation of Employment[a] (Distribution by Gender and Education Level).**

| Occupation of Employment | Proportion of Economic Activities[b] | Proportion of Males in Sector | Proportion of Total Males | Proportion of Total Females | Primary Education Completed Proportion | Ordinary Secondary Completed Proportion | Advanced Secondary Completed Proportion | Mean Monthly Earnings (in TZS) |
|---|---|---|---|---|---|---|---|---|
| | (1) | (2) | (3) | (4) | (5) | (6) | (7) | (8) |
| Legislators and Administrators | 0.01 | 0.72 | 0.01 | 0.01 | 0.89 | 0.59 | 0.53 | 954,550.60 |
| Professionals | 0.02 | 0.71 | 0.03 | 0.01 | 0.00 | 0.00 | 1.00 | 917,746.40 |
| Technicians and Associates | 0.03 | 0.49 | 0.03 | 0.04 | 0.82 | 0.88 | 0.56 | 479,540.80 |
| Clerks | 0.02 | 0.41 | 0.01 | 0.02 | 0.95 | 0.68 | 0.28 | 421,823.10 |
| Service and Sales Workers | 0.17 | 0.49 | 0.16 | 0.17 | 0.88 | 0.26 | 0.06 | 290,897.80 |
| Skilled Agricultural Workers | 0.42 | 0.48 | 0.38 | 0.45 | 0.62 | 0.06 | 0.01 | 78,414.59 |
| Craft and Related Workers | 0.10 | 0.81 | 0.15 | 0.04 | 0.88 | 0.18 | 0.03 | 249,822.80 |
| Plant and Machine Operators | 0.05 | 0.96 | 0.10 | 0.00 | 0.94 | 0.28 | 0.02 | 284,744.10 |
| Elementary Occupations[c] | 0.18 | 0.33 | 0.12 | 0.26 | 0.82 | 0.10 | 0.00 | 140,321.10 |
| Observations | 21,207 | 21,207 | 11,089 | 10,118 | 21,207 | 21,207 | 21,207 | 21,207 |

*Note*: Tanzania's 2014 Integrated Labor Force Survey. The sample includes males and females aged 15 years or more. [a]Categorization of occupation is based on Tanzania Standard Classification of Occupations (TASCO 1988) adapted from International Standard Classification of Occupations (ISCO 1988). [b]Economic activities denote a person's involvement in any work or activities for pay, profit, barter or home use in last twelve months (i.e. employment). Primary, Ordinary Secondary and Advanced Secondary refers to people who completed only Primary, Ordinary Secondary or Advanced Secondary schooling, respectively. [c]Elementary occupations include street vendors, shoe cleaning, domestic helpers, garbage collectors, agricultural, manufacturing, and transport labourers etc. Standard errors in parentheses. *** Significant at the 1 percent level. ** Significant at the 5 percent level.* Significant at the 10 percent level.

**Table A9. Returns to Education: By Occupation of Employment[a].**

| | Log of Monthly Earnings (TZS) | | | |
|---|---|---|---|---|
| | Tanzania mainland | | Dar es Salaam | |
| | (1) | (2) | (3) | (4) |
| Experience[b] | 0.05*** | 0.05*** | 0.05*** | 0.05*** |
| | (0.00) | (0.00) | (0.00) | (0.00) |
| Experience Squared | -0.00*** | -0.00*** | -0.00*** | -0.00*** |
| | (0.00) | (0.00) | (0.00) | (0.00) |
| Gender (1 if male) | 0.59*** | 0.61*** | 0.62*** | 0.47*** |
| | (0.04) | (0.04) | (0.09) | (0.09) |
| Primary (1 if schooling>=7) | 0.41*** | 0.27*** | 0.27*** | 0.21*** |
| | (0.04) | (0.04) | (0.07) | (0.07) |
| Ordinary Secondary (1 if schooling>=11) | 1.36*** | 0.93*** | 1.07*** | 0.78*** |
| | (0.05) | (0.05) | (0.08) | (0.08) |
| Advanced Secondary (1 if schooling>=13) | 2.38*** | 1.54*** | 2.16*** | 1.50*** |
| | (0.05) | (0.06) | (0.08) | (0.09) |
| Gender Dummy * Primary (1 if schooling>=7) | 0.05 | -0.01 | 0.07 | 0.07 |
| | (0.05) | (0.05) | (0.09) | (0.09) |
| Gender Dummy * Ordinary Secondary (1 if schooling>=11) | -0.17*** | -0.19*** | -0.15 | -0.03 |
| | (0.06) | (0.06) | (0.10) | (0.10) |
| Gender Dummy * Advanced Secondary (1 if schooling>=13) | -0.44*** | -0.47*** | -0.45*** | -0.31*** |
| | (0.07) | (0.06) | (0.11) | (0.11) |
| Legislators, Administrator, and Managers | | 1.07*** | | 1.13*** |
| | | (0.08) | | (0.10) |
| Professionals | | 1.12*** | | 1.00*** |
| | | (0.06) | | (0.07) |
| Technician and Related Professionals | | 0.73*** | | 0.58*** |
| | | (0.04) | | (0.06) |
| Clerks | | 0.75*** | | 0.73*** |
| | | (0.05) | | (0.06) |
| Service and Sales Workers | | 0.38*** | | 0.42*** |
| | | (0.03) | | (0.03) |
| Skilled Agricultural Workers | | -0.42*** | | 0.15** |
| | | (0.03) | | (0.07) |
| Craft and Related Workers | | 0.23*** | | 0.33*** |
| | | (0.03) | | (0.04) |
| Plant and Machine Operators | | 0.47*** | | 0.53*** |
| | | (0.03) | | (0.04) |
| Constant | 10.15*** | 10.24*** | 10.46*** | 10.40*** |
| | (0.04) | (0.04) | (0.07) | (0.07) |
| $R^2$ | 0.27 | 0.34 | 0.3 | 0.34 |
| Mean Log Monthly Earnings (TZS) | 11.71 | 11.71 | 11.71 | 11.71 |
| Observations | 16,817 | 16,817 | 7,256 | 7,256 |

*Note*: Tanzania's 2014 Integrated Labor Force Survey. The sample includes males and females aged 15 years or more. [a]Categorization of occupation is based on Tanzania Standard Classification of Occupations (TASCO 1988) adapted from International Standard Classification of Occupations (ISCO 1988). In regression analysis, our base occupation category is 'Elementary Occupations', which include street vendors, shoe cleaning, domestic helpers, garbage collectors, agricultural, manufacturing, and transport labourers etc. [b]Experience is calculated by taking the difference of one's age and one's schooling minus six years [this is the standard approach in the literature following Mincer (1974), Boissiere, Knight and Sabot (1985) and Lemieux (2006)]. Primary, Ordinary Secondary, and Advanced Secondary refers to people who completed only Primary, Ordinary Secondary or Advanced Secondary schooling, respectively. Standard errors in parentheses. *** Significant at the 1 percent level. ** Significant at the 5 percent level.* Significant at the 10 percent level.

**Table A10. Returns to Education: By Occupation of Employment[a].**

| | Log of Monthly Earnings (TZS) | | | | | |
|---|---|---|---|---|---|---|
| | Tanzania mainland | | | Dar es Salaam | | |
| | (1) | (2) | (3) | (4) | (5) | (6) |
| Experience[b] | 0.05*** | 0.05*** | 0.05*** | 0.05*** | 0.05*** | 0.05*** |
| | (0.00) | (0.00) | (0.00) | (0.00) | (0.00) | (0.00) |
| Experience Squared | -0.00*** | -0.00*** | -0.00*** | -0.00*** | -0.00*** | -0.00*** |
| | (0.00) | (0.00) | (0.00) | (0.00) | (0.00) | (0.00) |
| Gender (1 if male) | 0.53*** | 0.53*** | 0.61*** | 0.45*** | 0.46*** | 0.56*** |
| | (0.02) | (0.02) | (0.03) | (0.03) | (0.03) | (0.04) |
| Years of Schooling | 0.07*** | 0.07*** | 0.07*** | 0.08*** | 0.06*** | 0.06*** |
| | (0.00) | (0.01) | (0.01) | (0.00) | (0.01) | (0.01) |
| Legislators, Administrator, and Managers | 1.13*** | 0.12 | 0.11 | 1.21*** | 0.05 | 0.02 |
| | (0.08) | (0.28) | (0.27) | (0.09) | (0.36) | (0.36) |
| Professionals | 1.19*** | -2.81* | -3.60** | 1.09*** | -5.43** | -5.58** |
| | (0.05) | (1.65) | (1.48) | (0.06) | (2.36) | (2.25) |
| Technician and Related Professionals | 0.86*** | 0.31 | 0.27 | 0.66*** | -0.02 | -0.07 |
| | (0.04) | (0.20) | (0.19) | (0.05) | (0.30) | (0.30) |
| Clerks | 0.85*** | 0.29** | 0.26* | 0.79*** | 0.15 | 0.12 |
| | (0.05) | (0.14) | (0.14) | (0.06) | (0.19) | (0.18) |
| Service and Sales Workers | 0.40*** | 0.20*** | 0.18*** | 0.43*** | 0.20*** | 0.17*** |
| | (0.03) | (0.06) | (0.06) | (0.03) | (0.08) | (0.08) |
| Skilled Agricultural Workers | -0.38*** | -0.20*** | -0.22*** | 0.19** | 0.19 | 0.23* |
| | (0.03) | (0.05) | (0.05) | (0.07) | (0.14) | (0.14) |
| Craft and Related Workers | 0.24*** | 0.20*** | 0.17*** | 0.34*** | 0.27*** | 0.21** |
| | (0.03) | (0.07) | (0.07) | (0.04) | (0.10) | (0.10) |
| Plant and Machine Operators | 0.49*** | 0.58*** | 0.53*** | 0.55*** | 0.70*** | 0.63*** |
| | (0.03) | (0.07) | (0.08) | (0.04) | (0.09) | (0.09) |
| Legislators*Schooling | | 0.06*** | 0.09*** | | 0.08*** | 0.09*** |
| | | (0.02) | (0.02) | | (0.02) | (0.02) |
| Professionals*Schooling | | 0.20** | 0.26*** | | 0.34*** | 0.36*** |
| | | (0.08) | (0.07) | | (0.12) | (0.11) |
| Technician and Related Professionals*Schooling | | 0.03*** | 0.05*** | | 0.05*** | 0.06*** |
| | | (0.01) | (0.01) | | (0.02) | (0.02) |

| | Log of Monthly Earnings (TZS) | | | | | |
| --- | --- | --- | --- | --- | --- | --- |
| | Tanzania mainland | | | Dar es Salaam | | |
| | (1) | (2) | (3) | (4) | (5) | (6) |
| Clerks*Schooling | | 0.04*** | 0.05*** | | 0.05*** | 0.06*** |
| | | (0.01) | (0.01) | | (0.01) | (0.01) |
| Service and Sales Workers*Schooling | | 0.02*** | 0.03*** | | 0.03*** | 0.03*** |
| | | (0.01) | (0.01) | | (0.01) | (0.01) |
| Skilled Agricultural Workers*Schooling | | -0.03*** | -0.02*** | | 0.00 | 0.02 |
| | | (0.01) | (0.01) | | (0.02) | (0.02) |
| Craft and Related Workers*Schooling | | 0.00 | -0.01 | | 0.01 | 0.00 |
| | | (0.01) | (0.01) | | (0.01) | (0.01) |
| Plant and Machine Operators*Schooling | | -0.01 | 0.00 | | -0.01 | 0.00 |
| | | (0.01) | (0.01) | | (0.01) | (0.01) |
| Gender*Legislators*Schooling | | | -0.03*** | | | -0.02** |
| | | | (0.01) | | | (0.01) |
| Gender*Professionals*Schooling | | | -0.03*** | | | -0.03*** |
| | | | (0.00) | | | (0.00) |
| Gender*Technicians*Schooling | | | -0.03*** | | | -0.02*** |
| | | | (0.00) | | | (0.00) |
| Gender*Clerks*Schooling | | | -0.02*** | | | -0.02*** |
| | | | (0.01) | | | (0.01) |
| Gender*Service and Sales Workers*Schooling | | | 0.00 | | | -0.01 |
| | | | (0.00) | | | (0.01) |
| Gender*Skilled Agricultural Workers*Schooling | | | -0.01 | | | -0.06*** |
| | | | (0.01) | | | (0.02) |
| Gender*Craft and Related Workers*Schooling | | | 0.02*** | | | 0.01 |
| | | | (0.01) | | | (0.01) |
| Gender*Plant and Machine Operators*Schooling | | | -0.01 | | | -0.01 |
| | | | (0.01) | | | (0.01) |
| Constant | 10.05*** | 10.02*** | 10.00*** | 10.07*** | 10.18*** | 10.16*** |
| | (0.04) | (0.05) | (0.05) | (0.05) | (0.07) | (0.07) |
| $R^2$ | 0.34 | 0.34 | 0.34 | 0.34 | 0.34 | 0.35 |
| Mean Log Monthly Earnings (TZS) | 11.71 | 11.71 | 11.71 | 11.71 | 11.71 | 11.71 |
| Observations | 16,817 | 16,817 | 16,817 | 7,256 | 7,256 | 7,256 |

*Note*: Tanzania's 2014 Integrated Labor Force Survey. The sample includes males and females aged 15 years or more. [a]Categorization of occupations is based on Tanzania Standard Classification of Occupations (TASCO 1988) adapted from International Standard Classification of Occupations (ISCO 1988). The reference category is 'Elementary Occupations', which include street vendors, shoe cleaning, domestic helpers, garbage collectors, agricultural, manufacturing, and transport labourers etc. [b]Experience is calculated by taking the difference of one's age and one's schooling minus six years [this is the standard approach in the literature following Mincer (1974), Boissiere, Knight and Sabot (1985) and Lemieux (2006)]. Standard errors in parentheses. *** Significant at the 1 percent level. ** Significant at the 5 percent level. * Significant at the 10 percent level.